\documentclass[a4paper,11pt]{article}
\usepackage{jinstpub} 
\usepackage{lineno}
\usepackage{soul}
\usepackage{multirow} 
\usepackage{adjustbox}
\usepackage{graphicx}
\usepackage{siunitx}
\usepackage[english]{babel}

\title{\boldmath Resistive Fine Granularity Micromegas: 
Characterization and Performance for Different Spark Protection Resistive
Schemes}

\author[a,b]{M. Alviggi,}
\author[c]{M. Biglietti,}
\author[d]{M. T. Camerlingo,}
\author[a,b]{M. Della Pietra,}
\author[a, e]{C. Di Donato,}
\author[c,f]{R. Di Nardo,}
\author[g]{S. Franchellucci,}
\author[a]{P. Iengo,}
\author[c]{M. Iodice,}
\author[c,f]{F. Petrucci,}
\author[a]{G. Sekhniaidze,}
\author[h]{M. Sessa}
\affiliation[a]{INFN sezione di Napoli, Napoli, Italy}
\affiliation[b]{Universit\`a di Napoli Federico II, Napoli, Italy}
\affiliation[c]{INFN sezione di Roma Tre, Roma, Italy}
\affiliation[d]{INFN sezione di Bari, Bari, Italy}
\affiliation[e]{Universit\`a di Napoli “Parthenope”, Napoli, Italy}
\affiliation[f]{Universit\`a Roma Tre, Roma, Italy}
\affiliation[g]{D\'epartement de Physique Nucl\'eaire et Corpusculaire, Universit\'e de Gen\'eve, Gen\'eve, Switzerland}
\affiliation[h]{INFN sezione di Roma Tor Vergata, Roma, Italy}

\emailAdd{michela.biglietti@cern.ch}

\abstract{
The aim of the presented work is the development   of single-stage amplification resistive
Micro Pattern Gas Detectors (MPGD) based on Micromegas technology with the following characteristics: ability to
efficiently operate up to 10 MHz/cm$^2$ counting rate; 
scalability to large areas; 
fine granularity readout with small pads of the order of mm$^2$; 
good spatial and time resolutions (below \SI{100}{\micro\metre} and 10 ns, respectively).
The miniaturization of the readout elements and the optimization of the spark protection system, 
as well as the stability and robustness under operation, are the primary challenges of the project.
Two families of resistive patterns were realized using different techniques: pad-patterned embedded resistors and double-layer of Diamond Like Carbon (DLC) structures foils. 
The main difference between them  is that for the embedded resistors the charge is evacuated through 
independent pads, for  double-layer DLC the resistive layers are continuous and uniform and the charge 
is evacuated through a network of dot-connections, several millimetres apart. 
Using the DLC technique, a medium-size detector with an active area of 400 cm$^2$ was recently built and tested, with the main results reported in this paper.
Additionally, a large module (\numproduct{50 x 40} cm$^2$ active area), suitable for tiling large systems in future experiments, has been  successfully realised and is currently undergoing testing and performance studies.

The characterization and performance studies of the detectors 
were conducted using radioactive sources and an X-rays generator, 
with the detectors operated with various gas mixtures. 
A comparison of the results obtained with different resistive layouts and configurations 
is provided, with a particular focus on the response under high-rate exposure. 
Key results on tracking and timing performance from test-beam data for the latest constructed 
medium-size detector are also presented.
}

\keywords{Gaseous Detector; MPGD; resistive MPGD; Micromegas; High-rate capability}

\begin{document}
\maketitle

\flushbottom

\section{Introduction}
\label{sec:intro}
\begin{figure}[htbp]
\centering
\includegraphics[width=1.\textwidth]{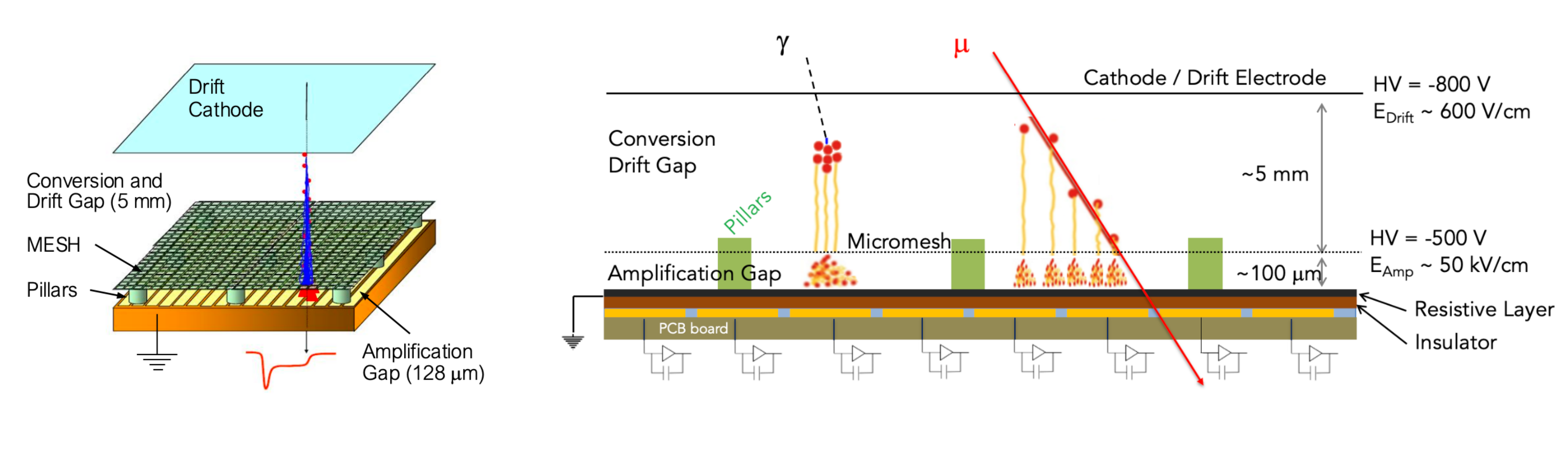}
\caption{Sketch of the Micromegas detector concept. The conversion/drift region and 
the amplification gaps are shown, electrically separated by a metallic micro-mesh structure 
which is held in position by approximately $\sim$\SI{100}{\um}  high pillars (left).
Principles of operations of a resistive Micromegas chamber  with 
reference operational parameters (right). The operating parameters are indicative and the
dimensions are not to scale.
\label{fig:micromegas}}
\end{figure}

Micromegas are single stage amplification gaseous detectors based on parallel plate electrode structure. 
The gas volume is electrically divided into two gaps by means of a stainless steel micro-mesh, kept in position by pillars: the conversion (or drift) gap, of a few millimetres, 
 and the amplification gap between the mesh and the anode plane, of about 0.1 mm, with the anode hosting the readout elements, usually micro-strips or pads. 
An electric field of few hundreds V/cm is applied in the drift region, while a more intense electric field with values of 40-50 kV/cm is supplied to the amplification gap. 
For high rate applications and/or intense flow of highly ionising particles, 
discharge effects are greatly mitigated with the implementation of a layer of resistive elements facing the amplification
gap~\cite{Alexopoulos-first}. 
This is, for example, the solution developed by the ATLAS experiment~\cite{ATLAS} for operations at a few tens kHz/cm$^2$~\cite{NSWTDR}.
Low resistivity strip Micromegas able to sustain high rates~\cite{Alexopoulos-resistive} are not suitable for scaling to large dimensions and suffer for large charging-up~\cite{Alexopoulos-first}. 
In figure~\ref{fig:micromegas}, the working principle of a Micromegas detector and a typical 
layout implementing a single resistive layer are reported. 
The signals are induced through capacitive coupling to the readout elements (strips or pads).

The main goal of the project is the development of resistive Micromegas detectors, aimed at stable operations under very high rates, up to 10 MHz/cm$^2$. 
The basic steps to achieve this goal are the optimisation of the spark protection resistive scheme 
and the miniaturization of the readout elements. In summary, the objectives of the project are:

\begin{itemize}
    \item A good rate capability up to 10s MHz/cm$^2$;
    \item Low occupancy, requiring fine granularity with readout elements (pixels/pads) with dimensions of few mm$^2$;
    \item Spatial resolution (depending on the application) of the order of \SI{100}{\micro\metre};
    \item Time resolution  (depending on the application) below 10 ns;
    \item Robustness and stable operation under high ionisation, 
    with limited spark rates, for long time;
    \item Working point with large margin before breakdown and instabilities onsets. 
\end{itemize}

Worldwide, other single-amplification-stage MPGDs, such 
as $\mu$-RWELL~\cite{bencivenni2} and $\mu$-RGroove~\cite{groove}, 
are actively undergoing R\&D efforts with similar objectives,
potentially enabling advancements across all related technologies. 
This constitutes a crucial step in detector R\&D, aimed at preparing MPGDs 
to meet the challenges posed by next-generation experiments and 
facilitating direct comparisons between different design approaches.
As of today, comparisons with alternative technologies highlight the 
resistive Micromegas developed in this work as particularly advantageous. 
These devices demonstrate excellent performance, exceptional robustness 
(characterized by stable operation at very high gains, providing 
a significant safety margin against discharges), and scalability.

Among the possible applications, there are the ATLAS very forward extension of muon tracking ~\cite{ATLAS-phase2}
(which is an option for future upgrades), 
as well as muon detectors or TPC at future accelerators. 
The current project is fully aligned with the ECFA roadmap for detector R\&D 
proposed by the European Strategy~\cite{ECFA}.
Specifically, within the realm of Gaseous Detectors (ECFA TF1), implemented through the 
DRD1 project~\cite{DRD1-Proposal}, 
our focus is on MPGD technologies proposed for the Muon systems and pre-showers/Calorimeters. 
Crucial developments in these areas must prioritize high rates and fine granularity.

\vspace{1.2\baselineskip}

Most of the R\&D phases and the characterization studies of the detectors 
and their performance,
have been carried out with small-size prototypes, which allowed us to optimise the 
 resistive configuration and the construction techniques. 
More recently we took a first step towards large area detectors, 
with dimensions compatible for use in next generation experiments. 
The first medium-size detector with fine readout granularity was built and fully tested.
The paper is organized as follows:  the design and construction features of the detectors are described in section~\ref{sec:detectors}.
The main characteristics and performance that distinguish them, are reported in 
section~\ref{sec:performance}.  In section~\ref{sec:Paddy400-testbeam},  the results obtained from  test beam data on recently produced medium-size detector are presented. Finally, section~\ref{sec:summary} provides a summary of the work presented  and future prospects.

We emphasize that this paper primarily addresses the performance of the detectors, 
particularly concerning fundamental characterization studies conducted 
in laboratories with radioactive sources.  
However, for the sake of completeness and to enhance the understanding of our accomplishments,
in section~\ref{sec:Paddy400-testbeam} 
we also present the key results obtained from  latest test-beam campaigns, 
with a focus on the most recently produced medium-size detector. 
A comprehensive examination of tracking performance, 
including detailed comparisons among different detectors conducted during test beams, 
will be the subject of a separate paper currently in preparation.


\section{Description of the detectors}
\label{sec:detectors}

During the initial years of the R\&D work, the goals were to optimize 
the structure and to explore the complementarity among different configurations 
to meet the requirements of the project. 
Most of these studies were conducted using small-scale prototypes. 
It's only in recent years that we have progressed to larger sizes, 
starting with a \numproduct{20 x 20} cm$^2$ detector and more recently initiating the construction 
of a \numproduct{50 x 40} cm$^2$ innovative resistive Micromegas.

\begin{figure}[htbp]
\centering
\includegraphics[width=.9\textwidth]{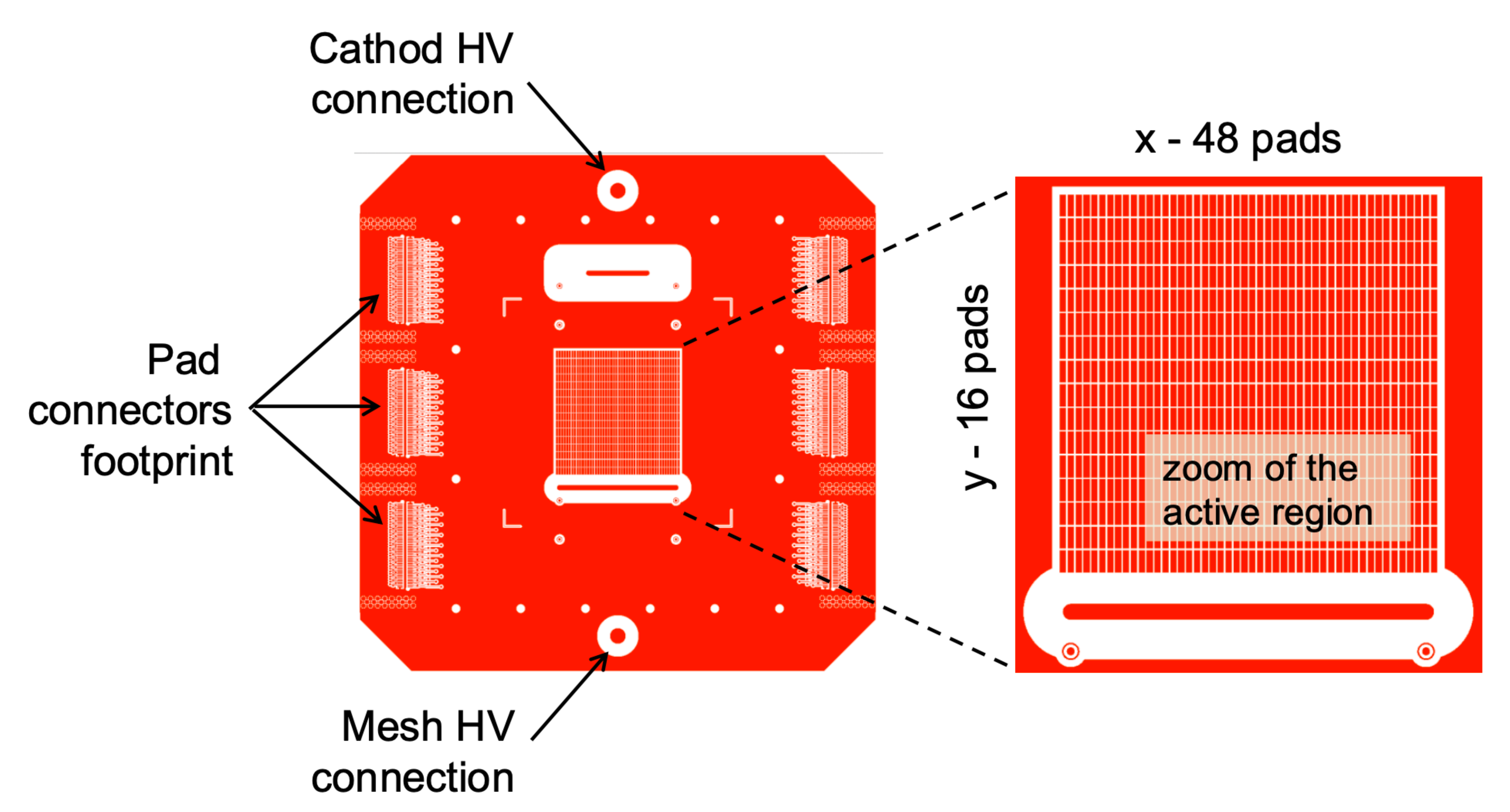}
\caption{Layout of the small-size prototype's anode plane 
with an expanded view of the pad structure.\label{fig:anode}}
\end{figure}

All small-size Micromegas prototypes presented in this paper consist of a similar anode plane, 
segmented with a matrix of \numproduct{48 x 16} readout pads. 
Each pad has a rectangular shape (0.8 $\times$ 2.8 mm$^2$) with a pitch of 1 and 3 mm in the two coordinates. 
The active surface is 4.8 $\times$ 4.8 cm$^2$ with a total number of 768 channels, 
routed off-detector for readout.
The layout of the anode plane can be seen in figure~\ref{fig:anode}.

\begin{figure}[htbp]
\centering
\includegraphics[width=.95\textwidth]{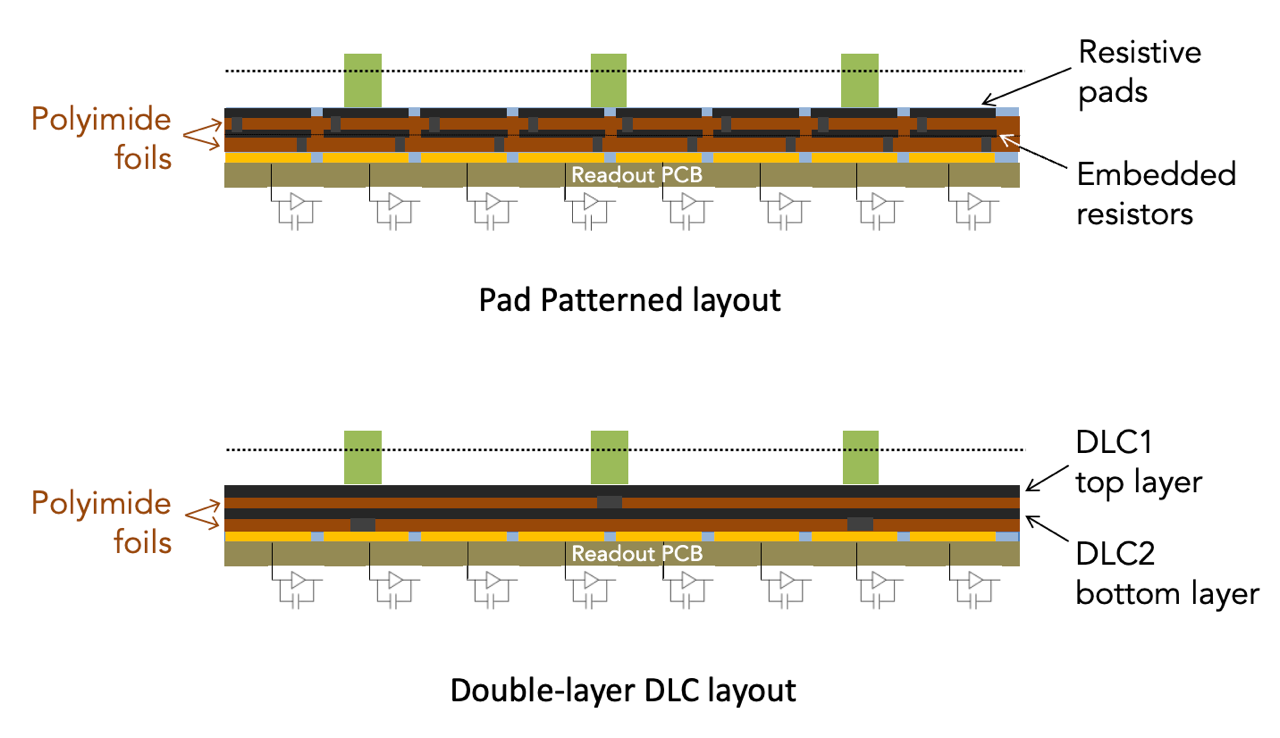}
\caption{Prototype configurations of the spark protection resistive layers:  pad-Patterned layout (top), double DLC layer (bottom).
\label{fig:resistiveschemes}}
\end{figure}

On top of the anode plane, different concepts and  configurations of the spark protection resistive layers 
have been implemented to evacuate the charge from the active area.
The two main concepts implemented are:

\begin{itemize}
    \item Pad-patterned (PAD-P) layout with embedded resistors (figure~\ref{fig:resistiveschemes}--top)
    based on resistive pads superimposed to the readout copper pads. The resistive and copper pads are separated by an insulating layer and electrically connected
  through an embedded resistor to evacuate the charge. 
    Each pad, in this configuration, is independent of the others. 
    Previous R\&D with this technique can be found in~\cite{Thibaud, Chefdeville1, Chefdeville2}.
    \item The second concept makes use of resistive foils
    based on f Diamond Like Carbon (DLC). 
    
The readout pads  are superimposed by a double-layer of DLC, deposited on an insulating film and incorporating a grid of staggered interconnecting vias for rapid charge evacuation
from the DLC layers to the pads, placed under the pillars as 
 sketched in figure~\ref{fig:resistiveschemes}--bottom.
Because of the uniform DLC layer, pads are not entirely independent as the charge can spread 
across multiple pads. 
It's worth noting that a double resistive layer, with staggered connection vias,
is  crucial to ensure the detector's robustness. Indeed, without this double-layer configuration, avalanches close to vias would encounter very low resistance, leading to instabilities.
Moreover,  this double-layer configuration provides nearly uniform resistance between the readout pads and the impact position of the electron avalanche,  regardless of the impact position of the electron avalanche.
Previous R\&D with this technique can be found in~\cite{Bencivenni}.
\end{itemize}
For each of these two main concepts, detectors with different configurations, 
materials and construction techniques have been built.


\subsection{The Pad-Patterned embedded resistors layout (PAD-P)}
\label{sec:padp}

We report the results of two detectors, also referred to as PAD-P2 and PAD-P3, 
where the pad-patterned layout was constructed with different methods and materials. 
PAD-P2, already extensively described in~\cite{paddypaper},
used the screen-printing technique to pattern both the top-layer resistive pads 
(facing the gas amplification gap) and the embedded resistor, 
both on \SI{50}{\micro\metre} thick polyimide foils.  
One drawback of this technique is that, for small pads,  embedded resistors with high enough resistance
are difficult to shape. 
This difficulty was overcome with the ``MIX'' technique, 
where the screen-printing is still used for the top-layer and the embedded resistors 
were obtained by patterning a DLC layer on the polyimide foil. 
Thanks to the higher resistivity of DLC (several M$\Omega/\square$)) with respect to the screen-printing paste (around 10 k$\Omega/\square$), 
this solution allows for easier patterning of small pads with high resistance~\cite{Rui-Instr2020}.
This is the solution adopted to build the PAD-P3 detector.

All these single-stage, small-pad resistive structures differ from earlier implementations for 
sampling calorimetry~\cite{Chefdeville1, Chefdeville2} due to the different pad size 
(30 times smaller) and construction techniques. 
They also differ from the developments for COMPASS~\cite{Thibaud}, where Micromegas 
are operated with a hybrid structure comprising a layer of GEM, 
given the impossibility of operating single-stage amplification non-resistive 
Micromegas with sufficient gain .

\subsection{The double-layer DLC layout (DLC)}
\label{sec:dlc}

\begin{figure}[htbp]
\centering
\includegraphics[width=.6\textwidth]{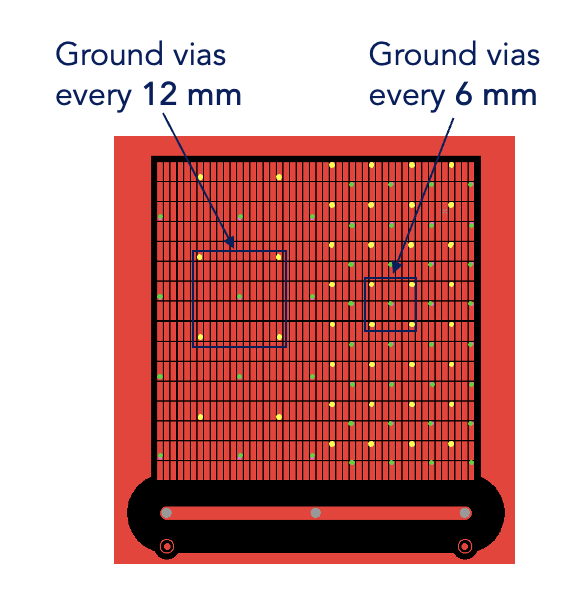}
\caption{Schematic view of the grounding vias for the double-layer DLC prototypes. 
The yellow dots represent the connections between the two DLC layers.
The green dots represent the connections between the bottom DLC layer and the pads 
(see figure~\ref{fig:resistiveschemes}--bottom).
\label{fig:vias}}
\end{figure}

The second scheme uses two uniform resistive layers of DLC, 
deposited by sputtering on polyimide foils and glued on the anode. 
The two resistive layers are interconnected with the readout pads with a network of conducting vias 
(filled with silver paste) with a few millimetres pitch to evacuate the charge, 
as sketched on figure~\ref{fig:resistiveschemes}--bottom. 
The first two detectors were built using the standard DLC technique, which involves 
drilling vias on DLC foils and manually filling them with silver paste. 
Different resistivity values were used.
The first detector has a resistivity in the range of 50-70 M$\Omega/\square$, referred to as DLC50, 
while the second detector has a resistivity of about 20 M$\Omega/\square$ and is named DLC20.
Both detectors feature an active plane divided into two halves, 
each with a different pitch of the conducting vias through the DLC layers: 
6 mm and 12 mm, respectively.  A schematic view  of the grounding vias  in this configuration is shown in figure~\ref{fig:vias}. 
In order to distinguish the two regions with 6 mm and 12 mm vias pitch, 
the suffixes “6-mm” and “12-mm” are added to the corresponding name.

\subsubsection{The DLC improved layout with the sequential build-up technique (DLC-SBU)}
\label{sec:sbu}


For the latest built prototypes, copper-coated DLC foils have been utilised to 
improve the construction technique, making use of the Sequential Build Up (SBU) 
method~\cite{Rui-Instr2020}, a process used in printed circuit board (PCB) industries to produce high-density interconnect boards.
These foils are etched leaving small copper circles (about 0.5 mm in diameter) in the position of the vias. The first layer is glued/pressed on the PCB, holes are drilled for the grounding evacuation vias, the plating technique (standard in PCB processing) is applied to connect the DLC with the pad. The same procedure is done for the second foil, plating the vias to connect the two DLC substrates. The final step is the bulking of the mesh ~\cite{Giomataris_Bulk}.

\begin{figure}[htbp]
\centering
\includegraphics[width=.8\textwidth]{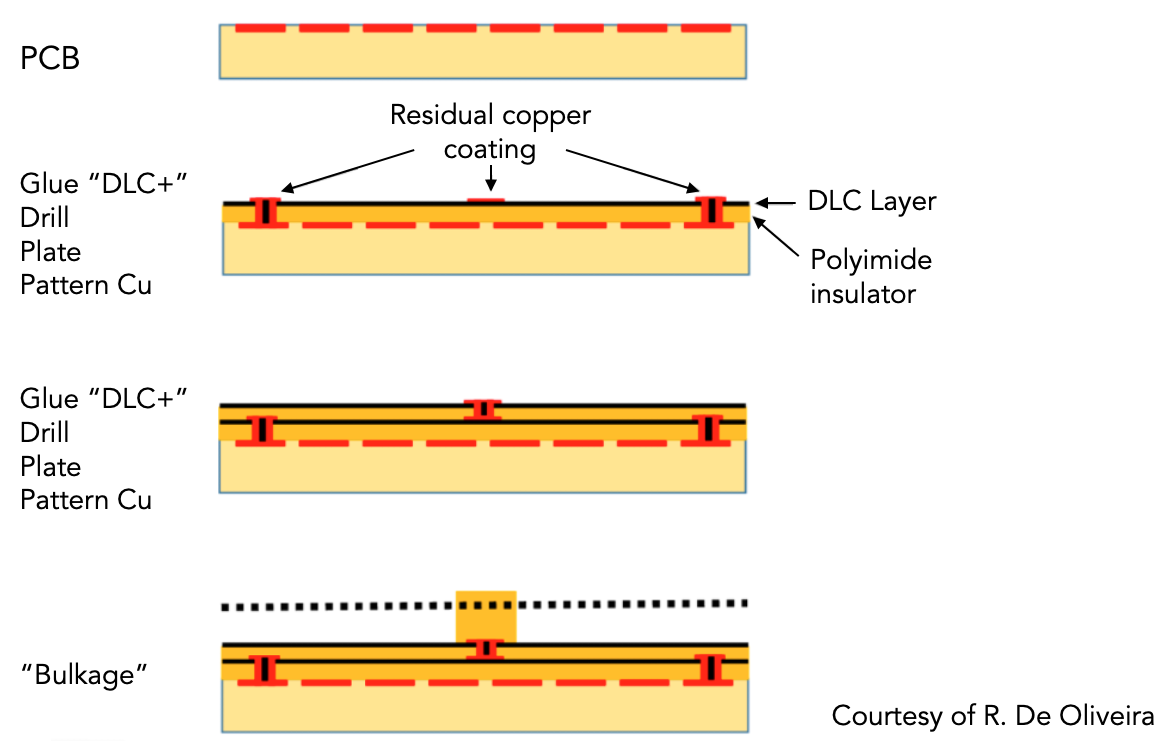}
\caption{Production steps of the double DLC foils resistive layout, implementing the Sequential 
Build-Up technique.
(Courtesy of Rui De Oliveira~\cite{Rui-Instr2020}). 
\label{fig:sbuproduction}}
\end{figure}

In figure~\ref{fig:sbuproduction}, the main steps to produce the SBU prototypes are reported. 
In this case, ``DLC+'' foils are used (DLC foil coated with \SI{5}{\um} thick Cu layer).

\begin{figure}[htbp]
\centering
\includegraphics[width=.98\textwidth]{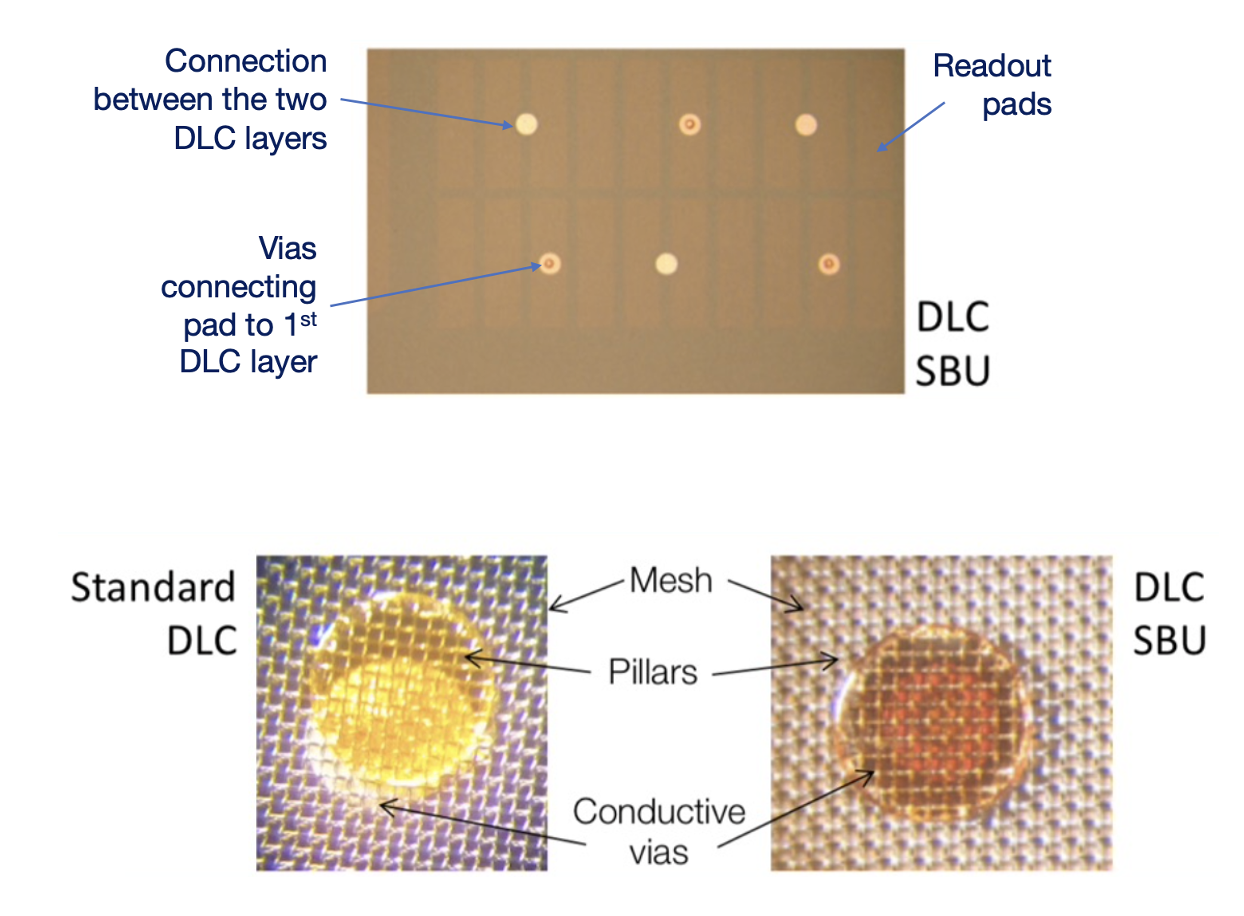}
\caption{Top: view of a small portion of the anode plane of a SBU prototype, after gluing of the first DLC layer, covering the readout pads (that can be seen in transparency). The copper circles to produce the conductive vias are visible. The smaller circles in some of them are the conductive vias between this DLC layer and the underlying pads; the “plain” copper circles will then host the staggered connections between this DLC layer and the top one, facing the gas gap. Bottom-left: example of pillar (uppermost yellow circle) in DLC20/50 prototypes, not well centred with the conductive vias.  The mesh layer can also be seen. Bottom-right: example of well centred pillar in SBU prototypes.
\label{fig:sbupillars}}
\end{figure}

The first advantage of this technique is that it allows to use the photo-lithographic 
process  
to precisely locate the conductive vias
(removing the copper everywhere except at the vias positions), make precise drilling,
and create the vias connections by plating. 
This technique allows for better alignment
of the vias below the pillars, thus preventing
sparks in those regions where the conductive vias can be misplaced and 
partially exposed to the gas gap. 
In figure~\ref{fig:sbupillars}, an example of a misalignment between the pillars and the vias in the standard DLC 
is shown, as compared with the pattern that was realised with the SBU technique.
Another advantage is that this technique is fully 
compatible with standard PCB processes, 
significantly facilitating the technological transfer of the production. 

With this technique the DLC-SBU prototype, whose results are reported in this paper, 
has been built, 
for which 
the configuration with the 6 mm pitch grounding vias is adopted in the full area. 
Despite a resistivity of $\sim$20 M$\Omega/\square$ was aimed for
for the DLC foils, the final measured values resulted to be around
35 M$\Omega/\square$ for the bottom resistive layer 
(closest to the anode pads) and around 5 M$\Omega/\square$ for the top layer (in the gas-gap region).
As a matter of fact, its behaviour is very close to the detector with both layers with 20 M$\Omega/\square$
(see section~\ref{sec:performance})~\footnote{Further investigation, through both
simulations and experiments, is needed to assess the behaviour of asymmetric configurations.}.


\subsection{The medium-size, 400 cm$^2$ fine granularity resistive Micromegas (Paddy400)}
\label{sec:paddy400}

\begin{figure}[htbp] 
\centering
\includegraphics[width=.9\textwidth]{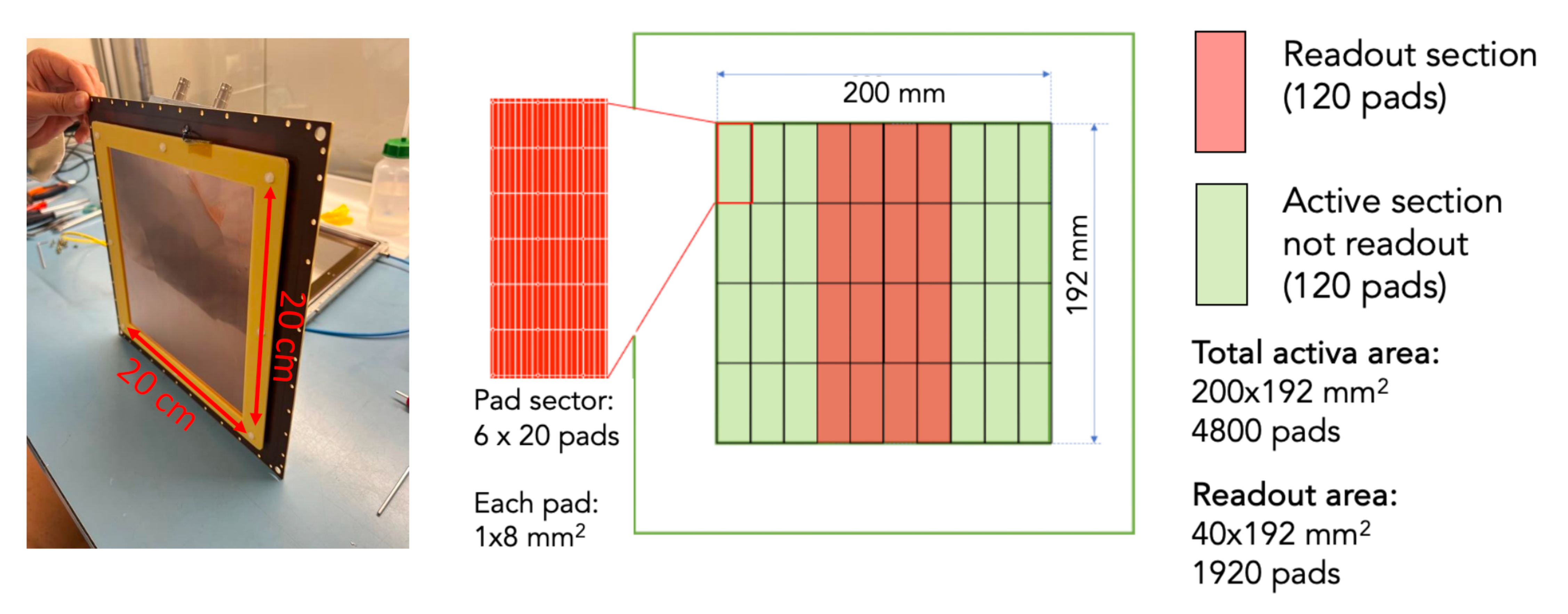}
\caption{Photo of a Paddy400 detector (left) and its layout of the readout pads (right).
\label{fig:Paddy400-layout1}}
\end{figure}

As an intermediate step towards the construction of improved resistive Micromegas for operation 
under high rates in large apparatuses, two similar medium-size detectors with an active area of 
400 cm$^2$ each, have been constructed (hereafter referred to as Paddy400-1 and Paddy400-2). 
The layout of the detectors is reported in figure~\ref{fig:Paddy400-layout1}. 

They implement the  double-layer DLC resistive configuration 
with grounding vias every 8 mm. 
Because of the  unavailability of copper-coated DLC foils, these detectors were built 
with the standard technique, (not SBU), 
with semi-automatic drilling and manual filling of the vias with silver paste.
In order to cope with possible misalignment between the conductive vias and the pillars, 
the pillar diameter was increased to about 0.7 mm, as compared with the 0.3 mm used
for SBU detectors. 

For both detectors the measured DLC resistivity is approximately 
25 M$\Omega/\square$ for the top/external DLC layer (close to the gas gap) and 35 M$\Omega/\square$ for the 
bottom/inner layer. They have an active area of \numproduct{200 x 192} mm$^2$, 
small pads readout elements of dimensions \numproduct{1 x 8} mm$^2$ 
(rectangular shape favouring one coordinate for high spatial resolution), for a total number of 4800 pads. 
Due to the high density of channels and mechanical constraints for the arrangement of the 
hybrid front-end boards on the back on the detectors, only a fraction of pads, 1920 out of 4800, are connected to 128-pin connectors. The other pads are grouped and can be connected 
 
\begin{figure}[htbp] 
\centering
\includegraphics[width=.7\textwidth]{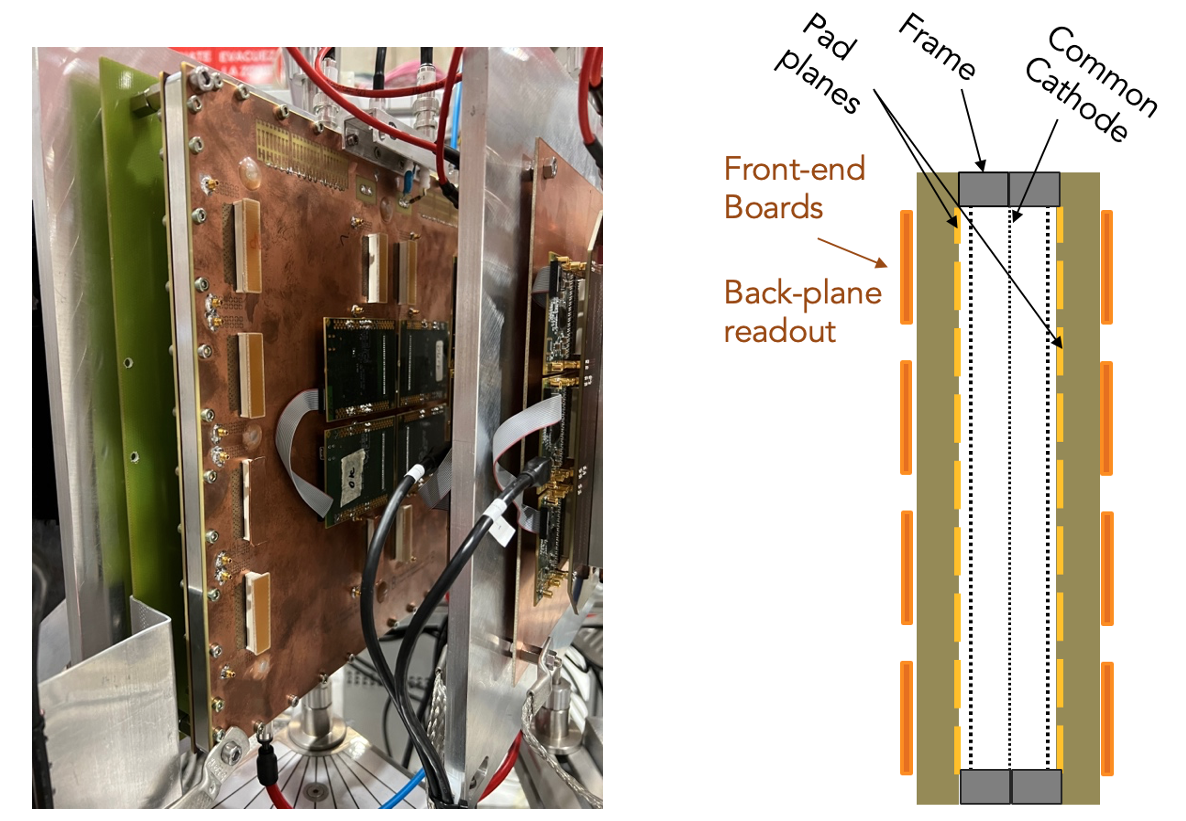}
\caption{Double Paddy-400 ``front-to-front'' detectors with a common cathode: photo of the detector (left), schematic view of the gap coupling (right).
\label{fig:Paddy400-layout2}}
\end{figure}

The design of the two detectors enables their ``front-to-front'' coupling, referred to as ``front-to-front'', 
wherein they share the same gas volume for the drift region, featuring a common cathode at the centre, 
as illustrated in the right part of figure \ref{fig:Paddy400-layout2}. 
This configuration facilitates the creation of a compact, low-material-budget, 
bi-layer Micromegas tracker. 
To ensure independent drift gaps and prevent possible charge migration from one side to another, 
the central cathode, suspended on the active area and smaller than the gas box, 
is constructed with a polyimide foil copper-coated on both sides.
This design allows gas circulation between the gaps from the sides of the active area.

\subsection{Summary of detector configurations and constructive parameters}
\label{sec:summary-table}

In table~\ref{tab:summary}, a summary of the configurations and the main constructive parameters of the detectors
under study is reported.
\begin{table}[h]
    \centering
     \begin{adjustbox}{max width=\textwidth}
    \begin{tabular}{|l|cccccc|}
    \hline
    \hline
        Detector Name &  \textbf{PAD-P2} & \textbf{PAD-P3} & \textbf{DLC20}  & \textbf{DLC50} & \textbf{DLC-SBU2} &  \textbf{Paddy400}\\
            \hline
            \hline  &  &  &  &  &  & \\[-1.5ex]   
                                & Pad Patterned & Pad Patterned & Double-layer & Double-layer  & Double-layer  & Double-layer\\
       Resistive configuration  & with embedded & with embedded & DLC & DLC & DLC & DLC \\
                                & resistors     & resistors     & & & & \\
            \hline &  &  &  &  &  & \\[-1.5ex]
        Construction technique & Screen-printing & "MIX" - Screen Printing & Standard & Standard & Sequential Build-up & Standard\\
                            &  & + Patterned DLC  &  &  &  &  \\
            \hline &  &  &  &  &  & \\[-1.5ex]
        Active area (cm$^2$) & \numproduct{4.8x4.8} & \numproduct{4.8x4.8} & \numproduct{4.8x4.8} & \numproduct{4.8x4.8} & \numproduct{4.8x4.8} & \numproduct{20x20}\\
            \hline &  &  &  &  &  & \\[-1.5ex]
        Readout pad pitch (x, y) [mm] &  (1, 3) & (1, 3) & (1, 3) & (1, 3) & (1, 3) & (1, 8) \\
            \hline &  &  &  &  &  & \\[-1.5ex]
        Average Resistance (M$\Omega$) & 5  & 19  & - & - & - & -\\
         (top-to-readout pad)) & & &  &  &  & \\
        \hline &  &  &  &  &  & \\[-1.5ex]
           
       DLC top / bottom layer    & - & - & 20 / 20 & 50 / 50  & 5 /35 & 25 /35 \\
       resistivity (M$\Omega/\square$) &  &  &  &  &  & \\
    \hline &  &  &  &  &  & \\[-1.5ex]
         Grounding vias pitch (mm)& - & - & 6 / 12 & 6 / 12 & 6 & 8 \\
             \hline &  &  &  &  &  & \\[-1.5ex]
         Pillars diagonal pitch (mm)&  4.2 & 4.2 & 4.2 & 4.2 & 4.2 & 5.7 \\ 
             \hline &  &  &  &  &  & \\[-1.5ex]
        Pillars diameter (mm) & 0.4  &  0.8 & 0.3 & 0.3 & 0.3 & 0.7\\ 
        
            \hline 
    \end{tabular}
    \end{adjustbox}
    \caption{Main features and constructive parameters of the prototypes under study.}
    \label{tab:summary}
\end{table}



For all the constructed detectors, a 400 lines per inch, woven stainless steel micro-mesh with \SI{18}{\um} diameter wires was used. 
The bulk Micromegas processing, through photo-imageable films, 
was adopted to encapsulate the mesh on the anode~\cite{Giomataris_Bulk}.
The resulting thickness of the amplification gap has a nominal value of \SI{128}{\um} 
(thickness of the photo-imageable film) which, after the bulk process, 
was observed to shrink to an average value of about \SI{100}{\um}. 
Variations in the order of few percent are expected in the construction of different detectors.
The pillars are typically spaced by several millimetres, and their diameters vary from 
0.3 to 0.8 mm, depending on the construction technique, as described above. All the detectors have the same drift gap of 5 mm.

As a concluding remark, we stress that the introduction of the SBU technique exploiting copper-coated DLC foils (now also available at CERN~\cite{bencivenni2}) significantly reduces the complexity of the production, allowing for the construction of the base of the detector at any PCB industries. The only remaining step, needing further work for the transfer of technology, is the bulking of the mesh. An effort is ongoing in this direction, whose description is beyond the scope of this paper.


\section{Characterization and performance}
\label{sec:performance}

The characterisation of the detectors has been done with converted X-rays,
either from $^{55}$Fe radioactive sources (of varying intensities) or from a 
copper target X-rays generator at the GDD Lab at CERN. 
The energy spectrum resulting from the conversion of $^{55}$Fe X-rays exhibits 
a prominent peak at 5.9 keV.
In argon-based mixtures, there is an additional secondary peak at 2.7 keV (argon-escape peak). 
On the other hand, the copper anode X-rays generator emits characteristic radiation at 8.04 keV,
accompanied by a substantial Bremsstrahlung contribution. 
We have used Garfield++~\cite{garfield, garfield2} to estimate the primary ionisations 
expected in the conversion gap, to be used as parameters for our studies.

Three gas mixtures were used:

\begin{itemize}
    \item Ar--CO$_2$ (93--7) was for a long time considered the nominal gas mixture for safe 
    operation under high irradiation, due to its stable and ageing-free features. 
    For example, it was initially assumed for operation of the Micromegas in the 
    ATLAS New Small Wheel (NSW)~\cite{NSWTDR}. Due to its low cost and simplicity to operate binary gas mixture, Ar--CO$_2$
    is often used in standard laboratory tests.
    \item Ar--CO$_2$--iC$_4$H$_{10}$ (93--5--2) Promoting and following the R\&D done with the ATLAS Collaboration, 
    the addition of 2\% of isobutane was tested to achieve 
    higher gain with higher stability~\cite{Proc_Stefano}. This addition shows no evidences of ageing effects up to 
    High-Luminosity LHC lifetime~\cite{ageing1, ageing2}. 
    To be noted that this argon based mixture with a content of isobutane less than 3\% is non-flammable. 
    \item Ar--CF$_4$--iC$_4$H$_{10}$ (88--10--2) The tetrafluoromethane was added to the 
    argon based gas mixture primarily to have a faster
    drift velocity for electrons, 
    which can drastically improve the time resolution~\cite{iodice-largesize} (see section~\ref{sec:Paddy400-testbeam}).
    Such a gas mixture has a significant Global Warming Potential (GWP) impact, 
    however, it is currently being considered only for R\&D performance studies.
\end{itemize}

The detectors were flushed with a gas flow of approximately 5 litres per hour, maintained at an overpressure of 2-3 mbar relative to the atmosphere.

A review of the results obtained with the detector using the pad-patterned resistive layer (PAD-P2) 
with Ar--CO$_2$ can be found in~\cite{paddypaper}. 
In summary, the main characteristics of this detector are the following: 
it is a very robust detector without any indication of early onset of discharges 
for avalanche multiplication gains up to few 10$^4$; 
the energy resolution is modest (at 5.9 keV as measured with $^{55}$Fe) in the range of 30-40\% FWHM; 
the spatial resolution in the precision coordinate (1 mm pitch of the pads) is about \SI{190}{\um};
it has a very good rate capability, up to X-rays irradiation with photon conversion 
rates as high as 100 MHz/cm$^2$.
Its behaviour with rates is, however, significantly affected by charging-up effects of the exposed 
insulated area at the edges of each pad.
A comparison of the performance of the Pad-Patterned structures with other prototypes is given in the next sections. 

\subsection{Experimental setup}
\label{sec:setup}

\begin{figure}[htbp]
\centering
\includegraphics[width=.98\textwidth]{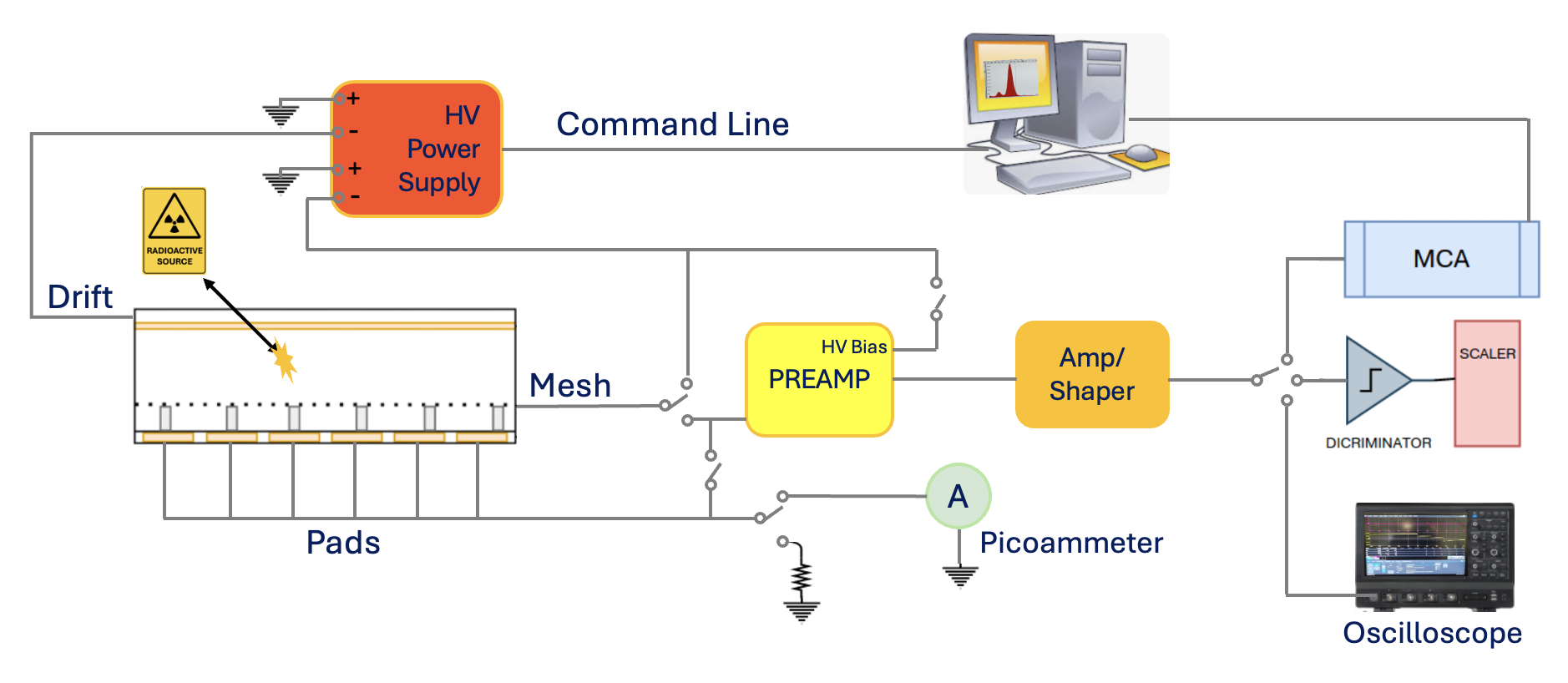}
\caption{Sketch of the typical experimental setup for laboratory characterisation of the 
detectors.
\label{fig:sketch-setup}}
\end{figure}

To comprehensively characterize the detector response to radioactive sources, 
several measurements are needed, including the number of detected particles, 
their frequency, signal intensity, energy spectrum of the source, 
and current and voltage values at the anode and cathode of the detector.
Figure~\ref{fig:sketch-setup} illustrates a typical setup featuring all 
configurations adopted for the results presented in this paper. 

The high voltage is provided by the power supply unit to the cathode 
(drift gap - HV$_{\rm drift}$). 
The amplification gap (HV$_{\rm amp}$) can be supplied to the mesh either using a 
direct connection or through the preamplifier HV bias. 
The direct connection, eliminating the preamplifier, 
is frequently employed for current mode measurements, 
particularly in cases of high-rate capability measurements, as elaborated 
in section~\ref{sec:ratecapability_method}.
This approach is used to eliminate any external resistance to ground,
which can be a primary factor leading to voltage drop in amplification, 
particularly under substantial detector current.
For the same reason, passive element filters are also removed during these 
specific measurements. 
Currents are measured and constantly monitored from the power supply and/or 
from the picoammeter (connected to either the mesh or the pads). 
A multichannel analyser (MCA) is employed for both spectrum analyses and rate measurements. 
The latter are also determined from a counter after discrimination. 
Pressure, temperature and relative humidity are monitored with sensors at the detector gas outlet 
(not reported in figure \ref{fig:sketch-setup}).

\subsection{Charging-up}
\label{sec:chargingup}

Very different charging-up behaviours have been observed for the PAD-P layout 
(embedded resistors with independent pads) and the layout using the uniform DLC foils, 
for what concerns the charging-up effects. 
As already reported in~\cite{paddypaper}, when the PAD-P detector is exposed to radiation, 
the detector current (i.e. the gain) initially reaches a maximum, and exponentially 
decreases with a time constant of the order of tens of seconds. 
The decrease reaches a plateau between 70-80\% of the initial value. 
It is worth mentioning that this effect is observed also irradiating the detector with relatively low rates,
of the order of kHz/cm$^2$, i.e., for detector current densities around 10 nA/cm$^2$. 
Such a behaviour is confirmed with other PAD-P detectors (namely PAD-P3) and has been observed 
also on resistive strip Micromegas~\cite{Alexopoulos-first}. 
In all these detectors, a substantial portion of insulation material is exposed within the amplification gap. 
This exposure arises from the pad-shaped resistive pattern, which leaves the insulator
polyimide substrate
uncovered around the pads. The fraction of the exposed insulator is of the order of 25$\%$ for 1$\times$3 mm$^2$ pads.
The charge accumulated during the irradiation reduces the intensity of the electric field in the gap,
decreasing the gain until a dynamic equilibrium is reached.  
This applies similarly to the space between the resistive strips for strip Micromegas.

Because of the different layouts of the detectors, not all of them are influenced equally by the charging-up. 
The detectors with the uniform DLC resistive layer are almost insensitive 
to this effect as the only insulator surface exposed in the amplification gap are the sides of the pillars. 
 
\begin{figure}[htbp]
\centering
\includegraphics[width=0.48\textwidth, trim={22cm 0 0 0},clip]{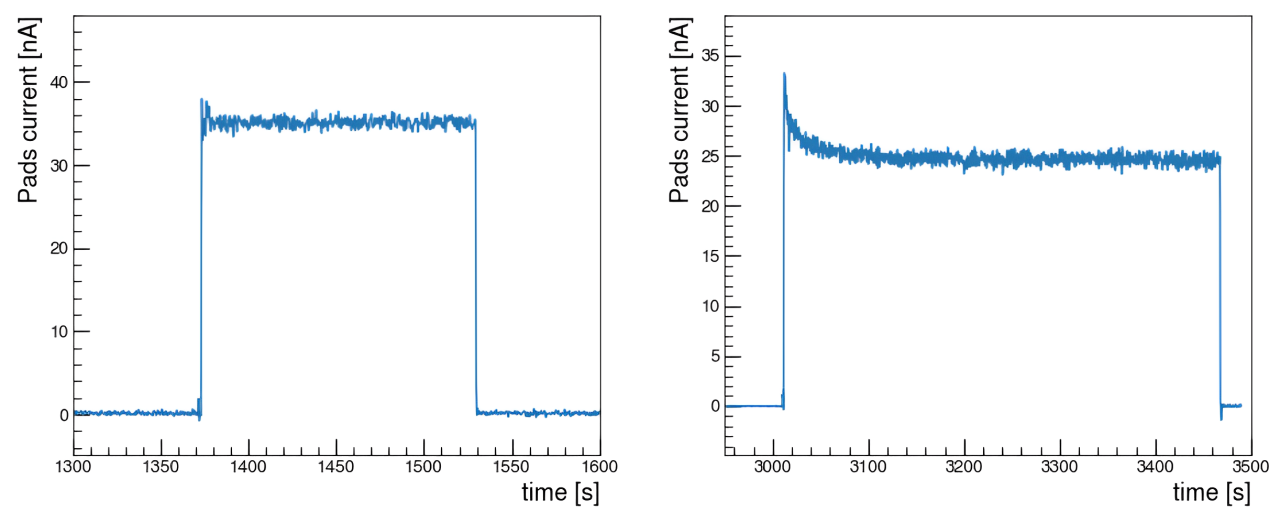}
\includegraphics[width=0.48\textwidth, trim={0 0 22cm 0},clip]{FIGURES/FIG_chargeUP_DLC_PADP.png}
\caption{Measured current as a function of time for a PAD-P (left) and a DLC prototype (right).
\label{fig:chargingupLowRate}}
\end{figure}

The different behaviour of PAD-P and DLC prototypes can be seen in figure~\ref{fig:chargingupLowRate}, where a PAD-P  (left) 
and a DLC (right) detectors were irradiated with a $^{55}$Fe source providing around 100 kHz hit rate,
switching it on and off over time, through a $\sim$1 cm$^2$ opening collimator. 
It can be seen that the double-layer DLC configuration doesn't show any dependence with time.

\begin{figure}[htbp]
\centering
\includegraphics[width=.65\textwidth]{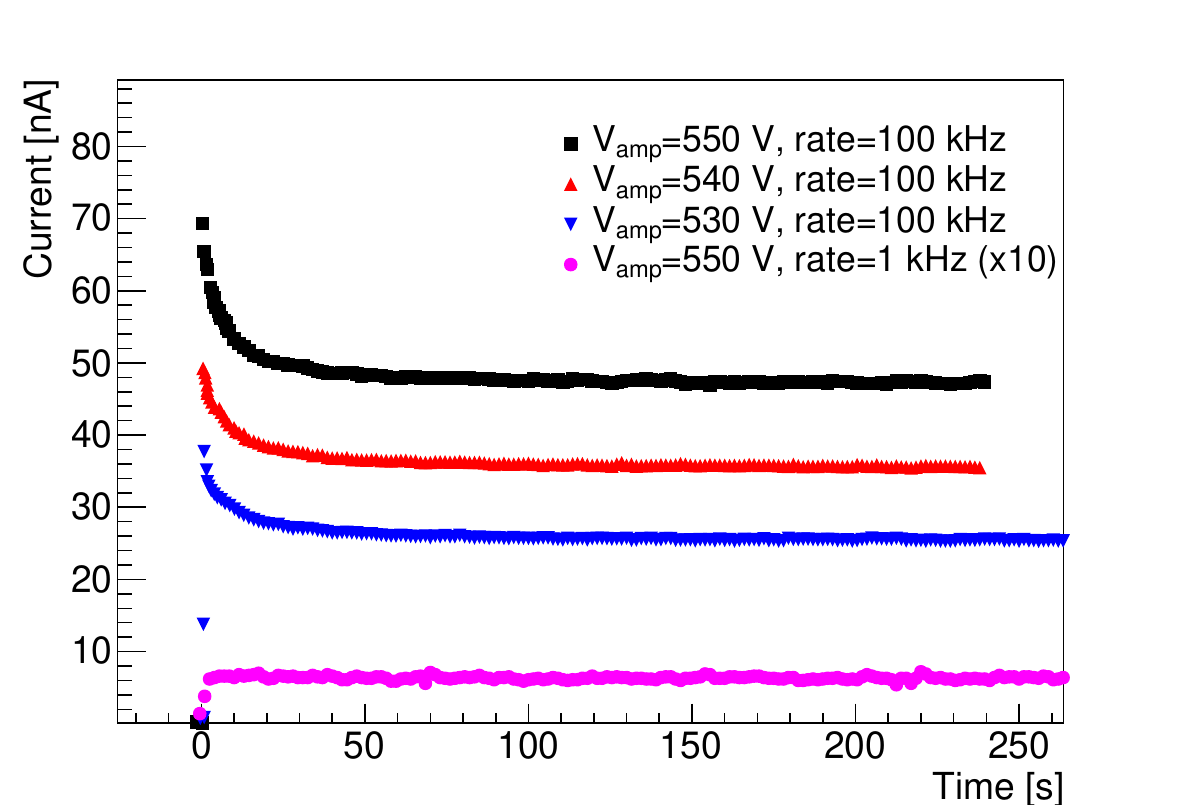}
\caption{Measured current in PAD-P2 prototype as a function of time, for different amplification voltages at a X-ray interaction rate of about 100 kHz and at a low rate (about 1 kHz) magnified by a factor 10. }
\label{fig:chargingupLowRate_PADP}
\end{figure}

\begin{figure}[htbp]
\centering
\includegraphics[width=.45\textwidth]{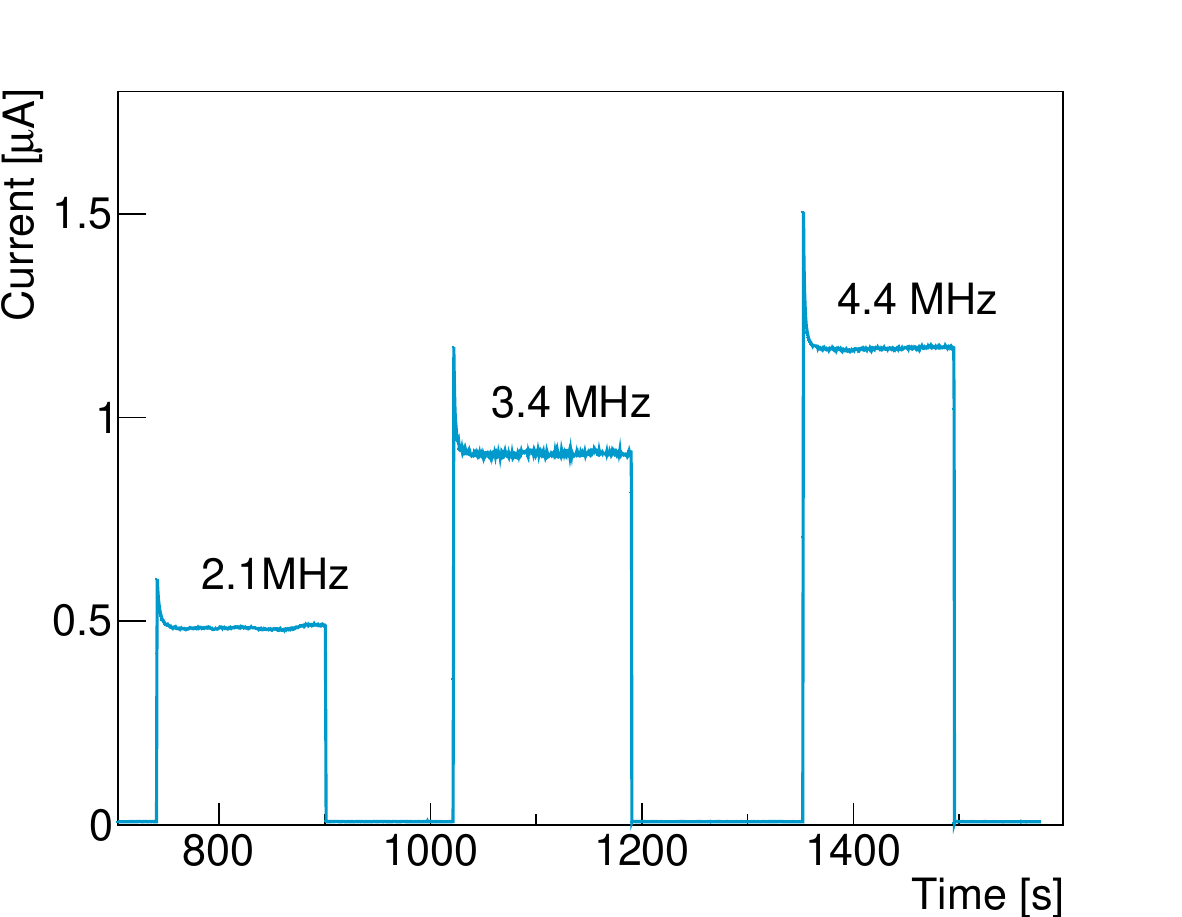}
\includegraphics[width=.47\textwidth]{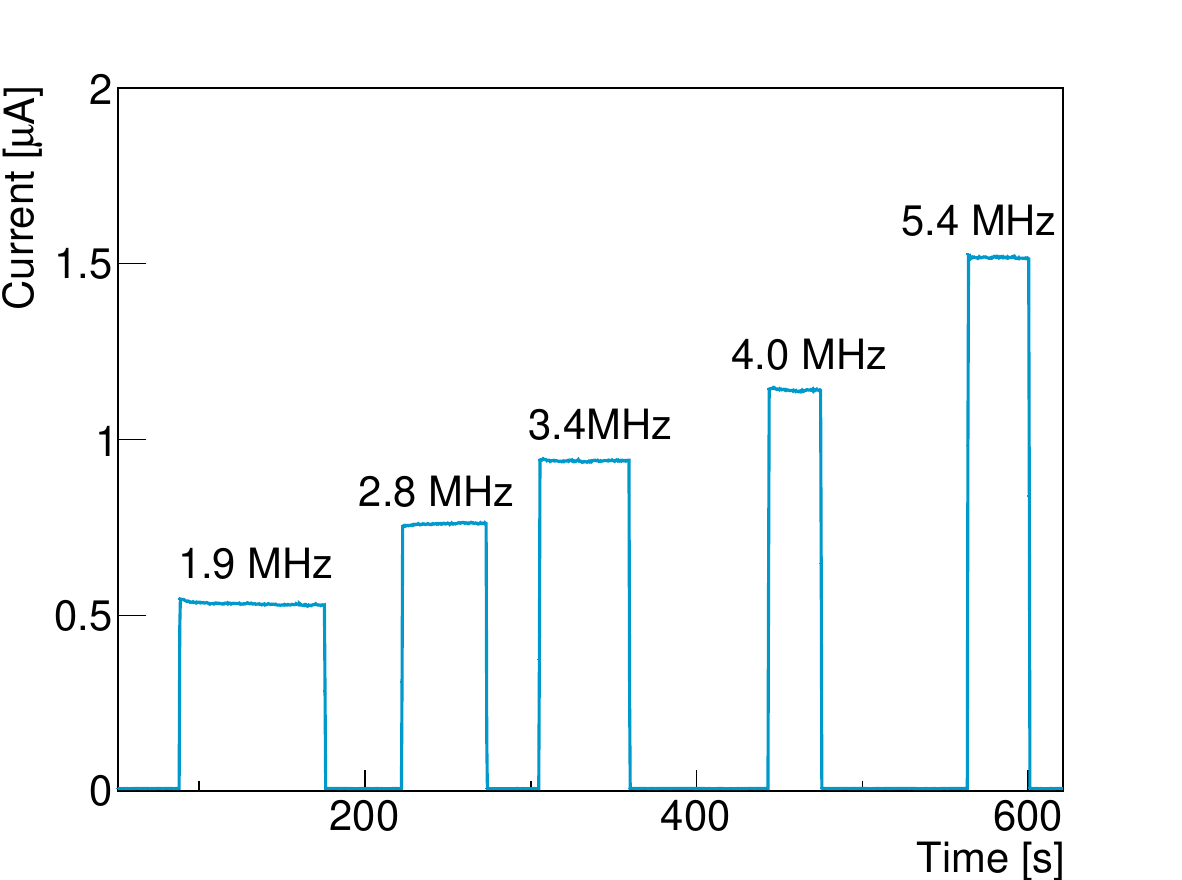}
\caption{Charging-up at high rates, in the region of MHz/cm$^2$, for a PAD-P type (left) and DLC type (right) prototype. 
Both operated at a gain of about $10^4$
\label{fig:chargingupHighRate}}
\end{figure}

In figure~\ref{fig:chargingupLowRate_PADP}, we report the results obtained with the PAD-P2 detector
exposed to a low and high intensity $^{55}$Fe sources ($\sim$1 kHz and $\sim$100 kHz hit rates, respectively), 
in the latter case, operating the detector at different gains. 
In the case of higher intensity source, a gain drop of about 30\% is observed.
Similar measurements were taken with a Cu X-ray generator to explore the regime of higher rates.
The results, reported in figure~\ref{fig:chargingupHighRate}, confirm the behaviour of a 
significant effect of charging-up for the pad patterned detector, 
while the one based on the double DLC layers don't show  any significant dependence with time
up the maximum measured
rate of 5.4 MHz/cm$^2$ (detector current at \SI{1}{\micro\ampere} over $\sim$1 cm$^2$ exposed area).

\clearpage

\subsection{Gain and maximum stability point}
\label{sec:gain}

The gain as a function of the amplification voltage is one of the basic 
measurements for the initial characterization of the detectors. 
To minimise (disentangle) possible effects of charging-up and pileup, 
the measurements are carried out at lowest possible rates, at the same time 
with a measurable detector current. 
A good compromise is found operating with a $^{55}$Fe source on 
spots of about 1 cm$^2$, 
giving a rate of $O$(10 kHz) converted X-rays and with currents in the range 
of  tens nA. 

Two methods have usually been adopted to measure the gain: 
current based and spectrum based.
The first one relies on the measurement of the detector current at the 
amplification stage $I_{\rm amp}$
(from the power supply or by a picoammeter - either from the mesh or from the pads) 
and from the measurement of the X-ray converted rate $R$. 
The effective gas multiplication factor (G) is determined using the formula 
\begin{equation}
 G = I_{\rm amp} / (R \times N_p \times e),   
\end{equation}
where $e$ represents the electron  charge
and $N_p$ the average number of electron-ion pairs produced from the interaction of one X-ray 
with the molecules of the gas. 
The average values of $N_p$ for the three different gas mixtures that we have used 
were estimated from simulations using Garfield~\footnote{N$_p$= $\{209, 211, 208\}$ for the Ar--CO2, Ar--CO2--isobutane, and Ar--CF4--isobutane gas mixture, respectively. 
}. 
The second method relies on the position of the main peak from the 
$^{55}$Fe source (at 5.9 keV) on a MCA calibrated scale.
Both methods have been validated to provide consistent results.

\begin{figure}[htbp]
\centering
\includegraphics[width=.45\textwidth]{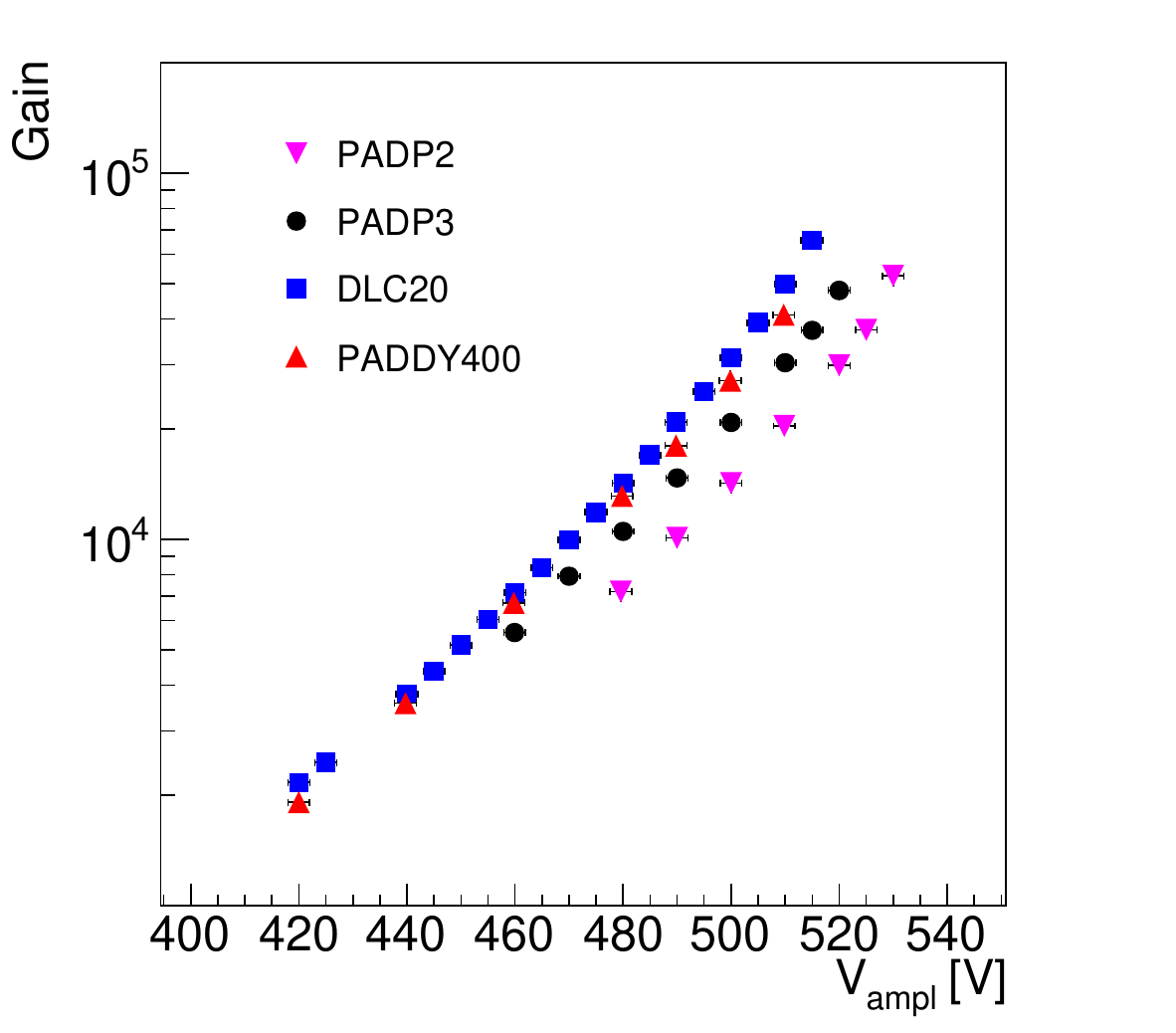}
\includegraphics[width=.45\textwidth]{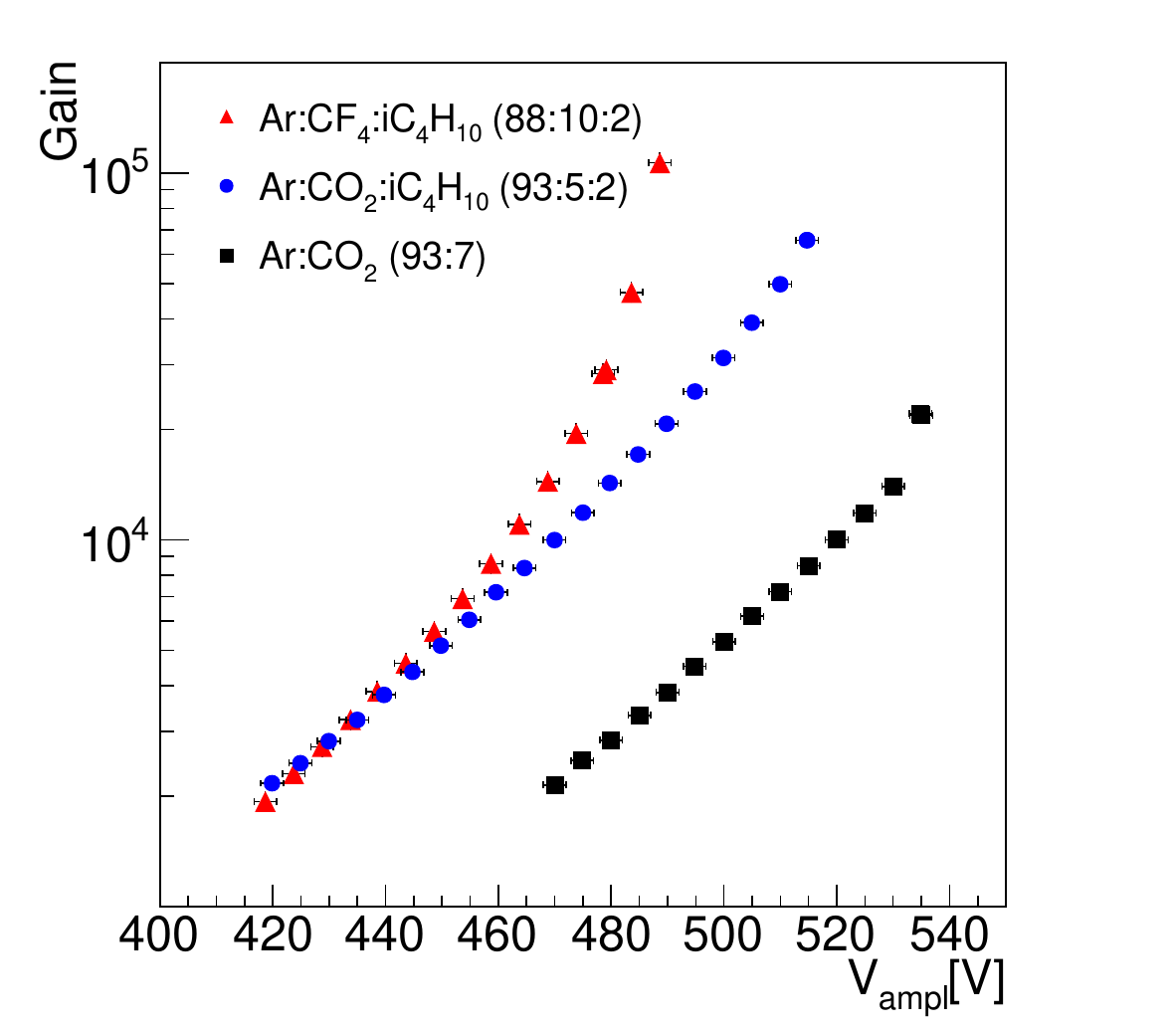}
\caption{Gain as a function of the amplification voltage for different detectors operated with Ar--CO$_2$--iC$_4$H$_{10}$ (93--5--2) gas mixture(left). Gain of DLC20 prototype with different gas mixtures
as a function of the amplification voltage (right).}
\label{fig:fig-gain}
\end{figure}

In figure~\ref{fig:fig-gain}--left, the gain curves as a function of the amplification voltage 
for two pad-patterned 
detectors  are reported along with two
double-layer DLC detectors (DLC20 and the large area, \numproduct{20x20} cm$^2$, 
Paddy400).
The measurements were obtained with the Ar--CO$_2$--iC$_4$H$_{10}$ (93--5--2) gas mixture.

Through extensive measurements of gain across various prototypes, valuable insights have been gained,
explaining the differences observed, despite the seemingly uniform construction conditions among prototypes,
including identical gas composition and (nominal) amplification gap dimensions. 
Notably, prototypes featuring pad-patterned designs with embedded resistors exhibit lower gains, 
a phenomenon attributable to both charging-up effects and less uniform electric field distribution.
Figure~\ref{fig:fig-gain}--right, shows the results for the DLC20 prototype 
operated with different gas mixtures, as a function of the amplification voltage.
For all measurements the drift electric field was kept at 600 V/cm. 
The measurements reported in the plot extend up to the last stable point before 
onset of discharges. The curves clearly show that the two mixtures 
containing 2\% of isobutane can significantly extend the stability range of the detector, 
up to $7\times 10^4$ and $10^5$ with Ar--CO$_2$--iC$_4$H$_{10}$ (93--5--2) 
and Ar--CF$_4$--iC$_4$H$_{10}$ (88--10--2), respectively. 
It can also be observed that 
the gas mixture containing CF$_4$ produces a sizable deviation from a simple exponential 
dependence as a function of the amplification voltage, possibly due to the increase of emission of 
secondary photons at high electric field.

It's important to note that small variations in amplification gap dimensions 
(ranging from a few to ten micrometers) are inevitable during production. 
According to simulations~\cite{gainVsGap}, 
even a minor variation of 5 micrometers 
in the amplification gap can lead to a substantial gain change of approximately 15--20\%. 
Moreover, it's crucial to acknowledge that the gain curves presented have not been adjusted 
for ambient pressure and temperature fluctuations. 
Such environmental factors can induce variations of up to 30\% in the observed gain values. 
 Nonetheless, the measurements reported in each plot are not affected by these fluctuations since they are acquired in stable conditions.

\subsection{Energy resolution}
\label{sec:energy_resol}

\begin{figure}[htbp]
\centering
\includegraphics[width=.44\textwidth]{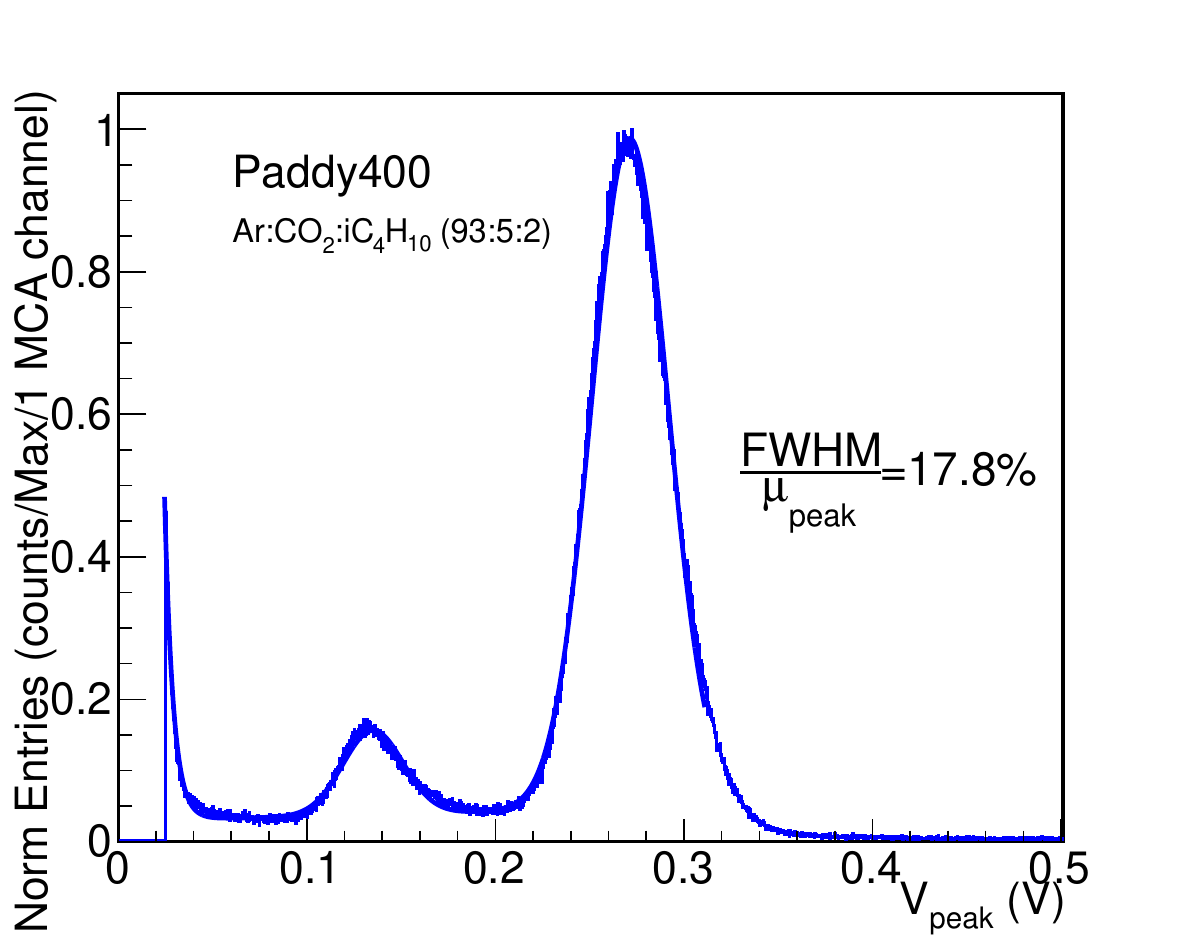}
\includegraphics[width=.45\textwidth]{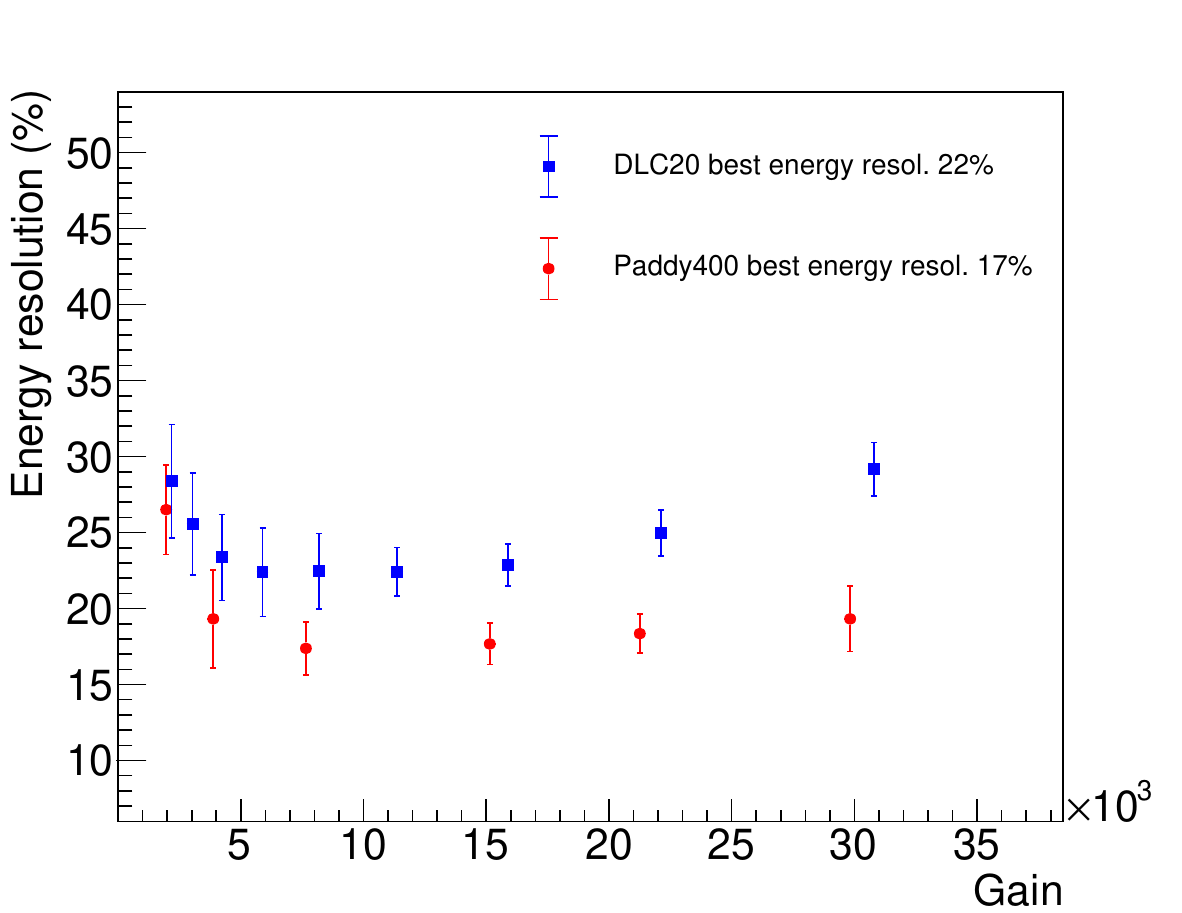}
\caption{$^{55}$Fe energy spectrum for Paddy400 prototype (left) and energy resolution of the small-size DLC20 prototype and medium-size Paddy400 prototype as a function of the gain (right). The minimum resolutions for both detectors are reported. 
\label{fig:energyresolution}}
\end{figure}

The energy resolution of the prototypes has been measured by fitting 
the energy spectrum of the incident radiation from a $^{55}$Fe source. 
The spectra are fitted assuming that the response is Gaussian, and that the low voltage region noise is described by an exponential distribution. An example is shown in figure~\ref{fig:energyresolution}--left, where
the $^{55}$Fe energy spectrum from Paddy-400, operated at gain of about 8000,
 is reported. The energy resolution is defined as the ratio between the Full Width at Half Maximum (FWHM) 
 and the position $\mu$ of the main photo peak from  Gaussian fit.

It's noteworthy that DLC detectors exhibit significantly superior energy resolution 
compared to PAD-P detectors: 20\% vs. 34\% at 5.9 keV. 
These values are representative of typical performance. 
The comparatively modest energy resolution observed in PAD-P detectors, 
extensively discussed in~\cite{paddypaper}, is attributed to the non-uniform structure of the 
amplification electric field arising from the resistive pad edge effects. 
Conversely, in the DLC configuration, the uniformity of the DLC foils 
without pad patterning ensures a more uniform electric field with smaller 
variations in amplification.

In figure~\ref{fig:energyresolution}--right, the dependence of the 
energy resolution at 5.9 keV as a function of the gain is reported 
for the DLC20 detector and for the  large-size \numproduct{20x20} cm$^2$ Paddy400 detector. 
Notably, for gains exceeding $\sim$4000, the best value of the energy 
resolution is achieved, in the range 17--22\% depending on the specific 
detector under test. At large gain, a worsening of the energy resolution is observed due to the increase of the avalanche size~\cite{sauli}.
The slight variations of a few percent observed in detectors with very
similar structures are not fully understood but could be related to local effects, like small irregularities in the amplification gap size, pillar size or positions with respect to the source, etc.


\subsection{Rate capability and ion backflow}
\label{sec:ratecapability}


The rate capability concept for a detector comprises two main different aspects: 
\emph{i)} its capability  to keep stable performance at low and high rates; 
\emph{ii)} the possibility to read out all strips/pads for each event without efficiency loss. 
While the first aspect can be attributed to the intrinsic characteristics of the detector, 
the second one strongly depends on the DAQ architecture, bandwidth, and the performance 
of the front-end and back-end electronics.

In this section we focus on the first aspect, by studying the detector's response to increasing fluxes of radiation, 
more specifically, measuring the gain stability and the gain drop occurring at high rates 
typically due to two different effects:  charging-up and voltage drop, the latter due to the resistive layer. 
Concerning the readout capability, however, we designed the detectors with fine granularity in order to minimise the occupancy.
Considering the size of 3 mm$^2$ pads, the rate per readout element for 
fluxes of 1 MHz/cm$^2$, is $\sim$30 kHz per readout channel, 
very well manageable by both the established APV~\cite{apv} and  new generation multichannel
(e.g. VMM~\cite{vmm}, Tiger~\cite{tiger}, VFAT3~\cite{vfat1, vfat2}, etc.) high density chips  .
Of course, the capability to handle these rates depends on the DAQ architecture, 
that should foresee fully parallel readout links according to the limits in the bandwidth.

\subsubsection{Description of the method and calibrations}
\label{sec:ratecapability_method}

To investigate the detector response, particularly their gain, across a broad spectrum of rates, they have been exposed to the copper target X-ray generator. The rates of X-rays can be regulated by varying the  X-ray filament current, $I_{Xray}$, in a range between 0 and 2.4 mA, with a precision of 0.001 mA. 
It is important  to clarify that the term "incident rate" 
 as used in the following context, refers to the measured events. 
 More precisely, it represents the convolution of the rate emitted by the source and the sensitivity of the detector.
The measurements were carried out in two steps, for low and for high rates:
 \begin{itemize}
     \item Step 1, low rate measurements : on a limited range of  $I_{Xray}$ and at low detector current $I_{det}$, with the pre-amplifier and MCA connected to the detector. In this step we measure all the parameters: $I_{Xray}$ (set), rate, and $I_{det}$. In this region we must find a linear dependence of $I_{det}$ and rates as a function of $I_{Xray}$. 
     \item Step 2, high rate measurements: the preamplifier is removed and rates cannot be directly measured. In this step we measure  $I_{Xray}$ (set), $I_{det}$ and we extrapolate the rates linearly from the measurements of rates vs $I_{Xray}$, done in step 1. The measurement in this case is extended to the full range of $I_{Xray}$; a loss of rate capability would manifest itself as a deviation from the linear behaviour of $I_{det}$ (and hence the gain) as a function of  $I_{Xray}$.
 \end{itemize}
 
\begin{figure}[htbp]
\centering
\includegraphics[width=.75\textwidth]{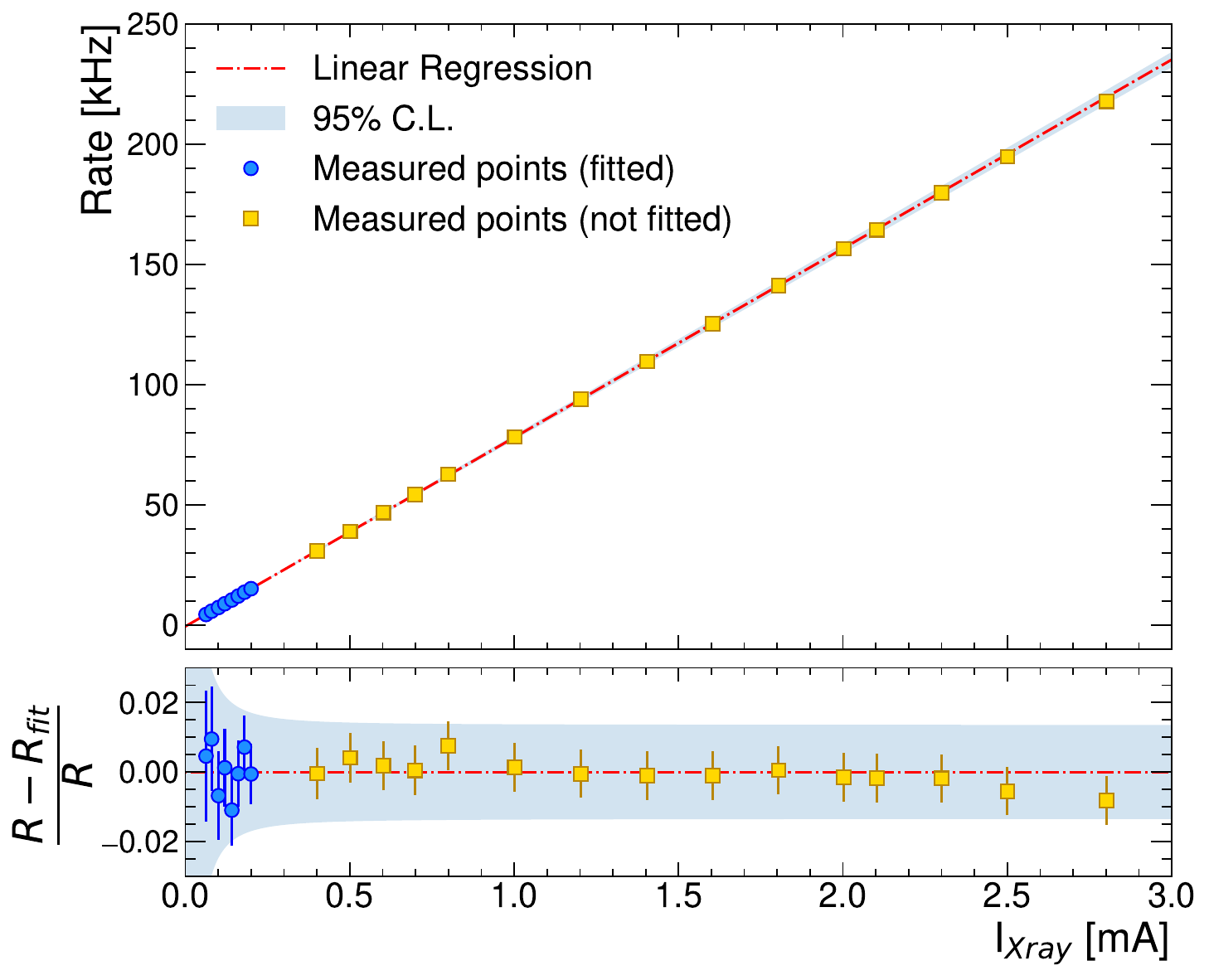}
\caption{
Measurements done with the MCA for checking the linearity between the current on the on X-ray filament (I$_{Xray}$) 
and rate of generated X-rays; 
the top plot shows the measured points together with the linear regression of the first 8 points, 
in the bottom one the difference between the fitted line and the experimental point is reported. 
All the points fall into the 95\% confidence interval of the extrapolation.
\label{fig:xraycalib}}
\end{figure}

 The second step assumes a linear response of the X-rays generator, in terms of rates vs $I_{Xray}$ in the full range. In order to validate this linearity, a calibration measurement was done using a copper absorber to keep X-rays rate low enough to be measurable (no pileup) and to ensure operation of the detector at full efficiency. A linear dependence between X-ray rates and the X-ray filament currents is observed, as shown in figure~\ref{fig:xraycalib}. A linear fit was performed only for the low current points and its extrapolation checked in the full range. Only the last points are below the extrapolated rate but in agreement with them within 2$\%$. 
As the linearity is preserved also without the copper absorber, the  extrapolation performed in step 2 is valid also at high values of the rates.

\subsubsection{Comparison between the Pad-Patterned and DLC layouts}
\label{sec:differencePADDLC}
With the described strategy, the rate capability was measured in an extended range, 
more than four order of magnitudes, of incident rates.

\begin{figure}[htbp]
\centering
\includegraphics[width=.8\textwidth]{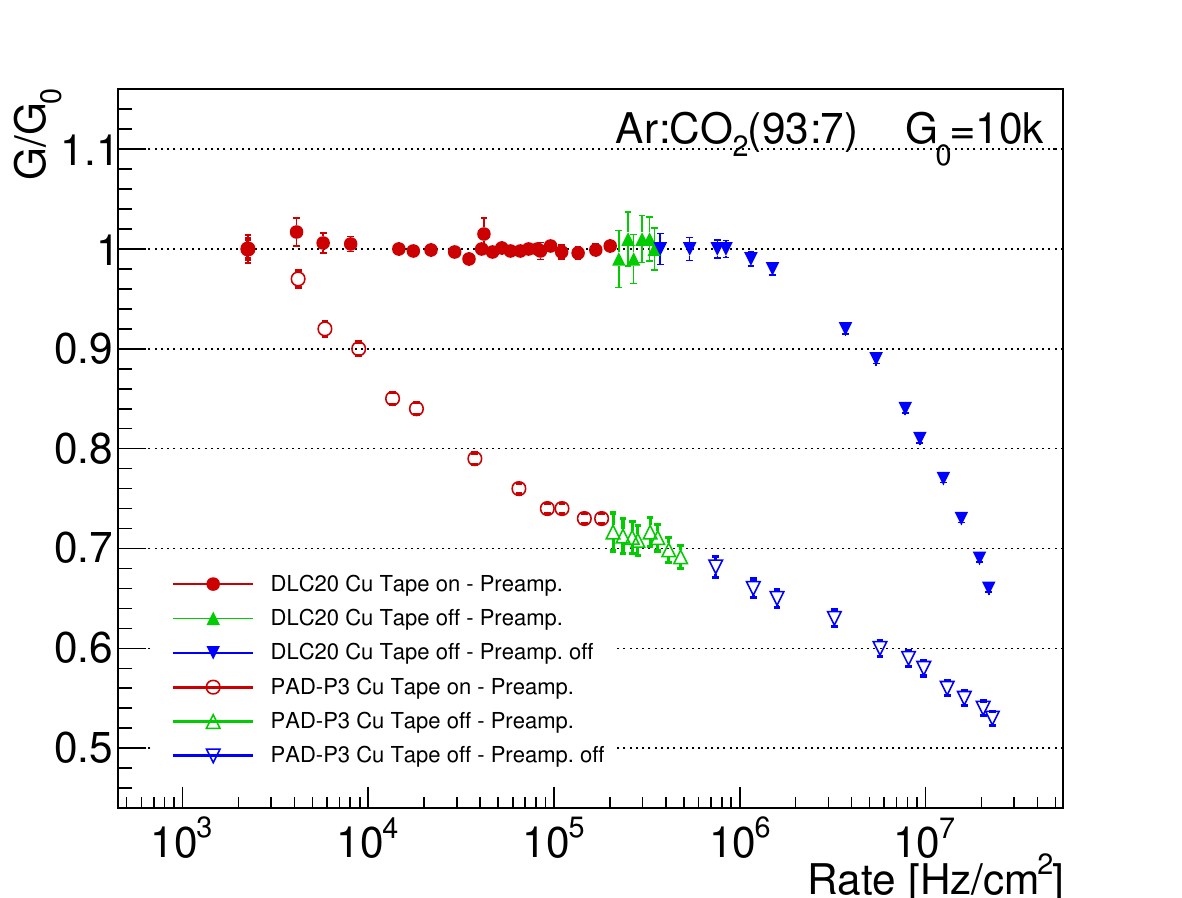}
\caption{Dependence of the gain normalised to its value at low rates 
for the pad-patterned PAD-P3 detector and double DLC layer 
DLC20 detector. 
\label{fig:ratecapability-padp-dlc}}
\end{figure}

It must be stressed that the ionization occurring with $\sim$8 keV photons from the 
X-ray generator is  around a factor of five higher than the ionisation of MIP particles traversing a 
drift (conversion) gap of 5 mm, and so, also the rates should be scaled-up by the same factor for a MIP-equivalent ionisation. 
From Garfield simulations in Ar--CO$_2$ (93--7), the number of e-ion pairs is $> 250$ from 8 keV photons and $\sim 50$ for MIPs. 
In figure~\ref{fig:ratecapability-padp-dlc} the gas gain, normalized to the reference value of $G_0$ = 10000, 
is reported up to 30 MHz/cm$^2$, as obtained with the PAD-P3 and DLC20 detectors with a gas mixture Ar--CO$_2$ 93--7. 
Data are taken with a shielding defining a uniform irradiated area in a circle with a diameter of 1 cm corresponding to an area of 0.79 cm$^2$. A thin copper tape  was used as absorber to explore a wider range of incident fluxes. At low irradiation level, it is possible to measure the incident rate directly from the preamplifier. To achieve  rates above 100 kHz/cm$^2$ the tapes are removed, and at higher currents the measurements are done employing the procedure described in section~\ref{sec:ratecapability_method}. 
We observe that the PAD-P3 prototype, due to the charging-up of the insulator surfaces, rapidly 
loses gain from low rates up to approximately 0.1 MHz/cm$^2$.
Subsequently, the gain reduction progresses more slowly as the pads exhibit behaviour 
like being nearly isolated from each other, 
and the ohmic voltage drop is only related to the small amount of current (increasing with rates) 
flowing through the resistance of a single pad.
On the other hand, the DLC type remains highly stable up to 1 MHz/cm$^2$, 
but its gain drops more rapidly at higher rate values.
Nevertheless, the overall outcome is noteworthy, our prototypes 
demonstrate excellent performance, sustaining high gain levels without discharges, 
even at rates exceeding 10 MHz/cm$^2$.

\subsubsection{Dependence on various parameters of the Double-Layer DLC layout}

\begin{figure}[htbp]
\centering
\includegraphics[width=.46\textwidth]{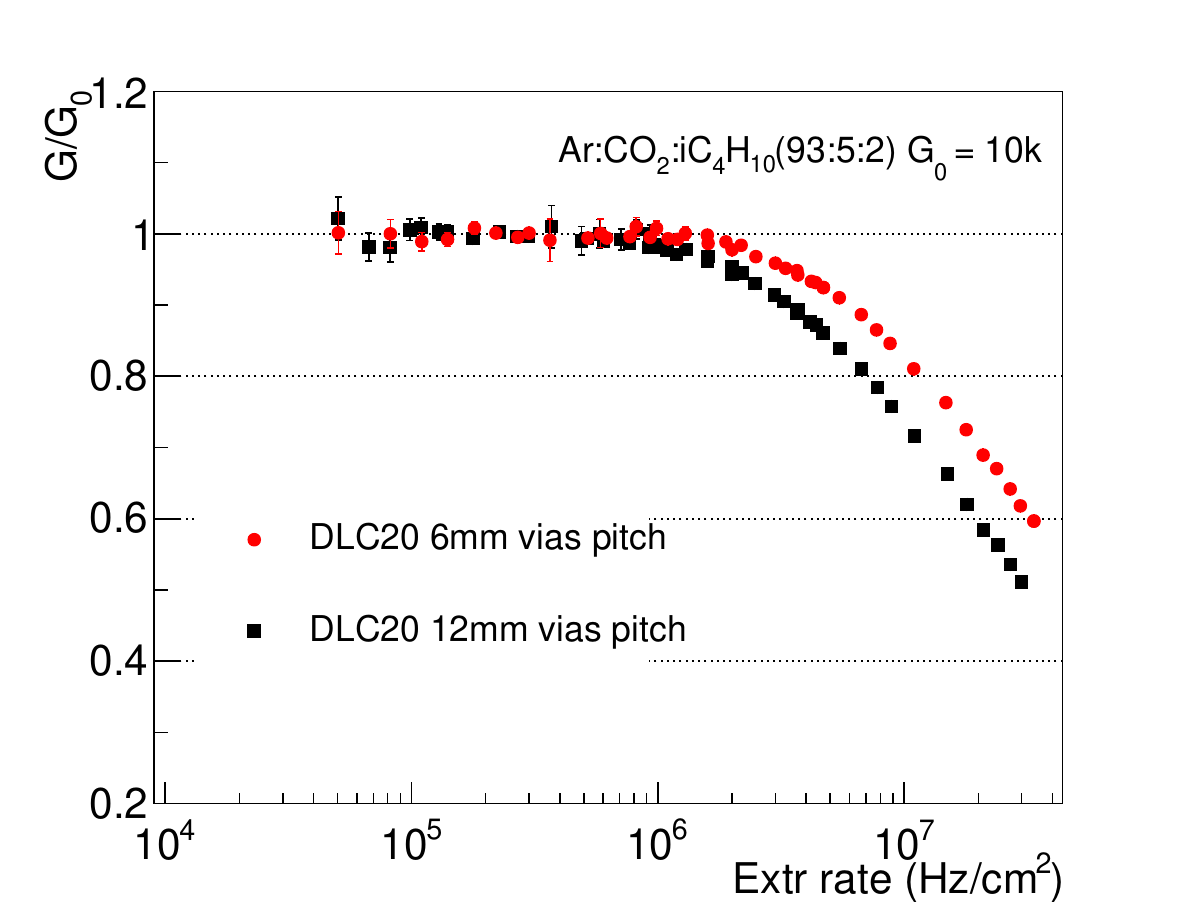}
\includegraphics[width=.46\textwidth]{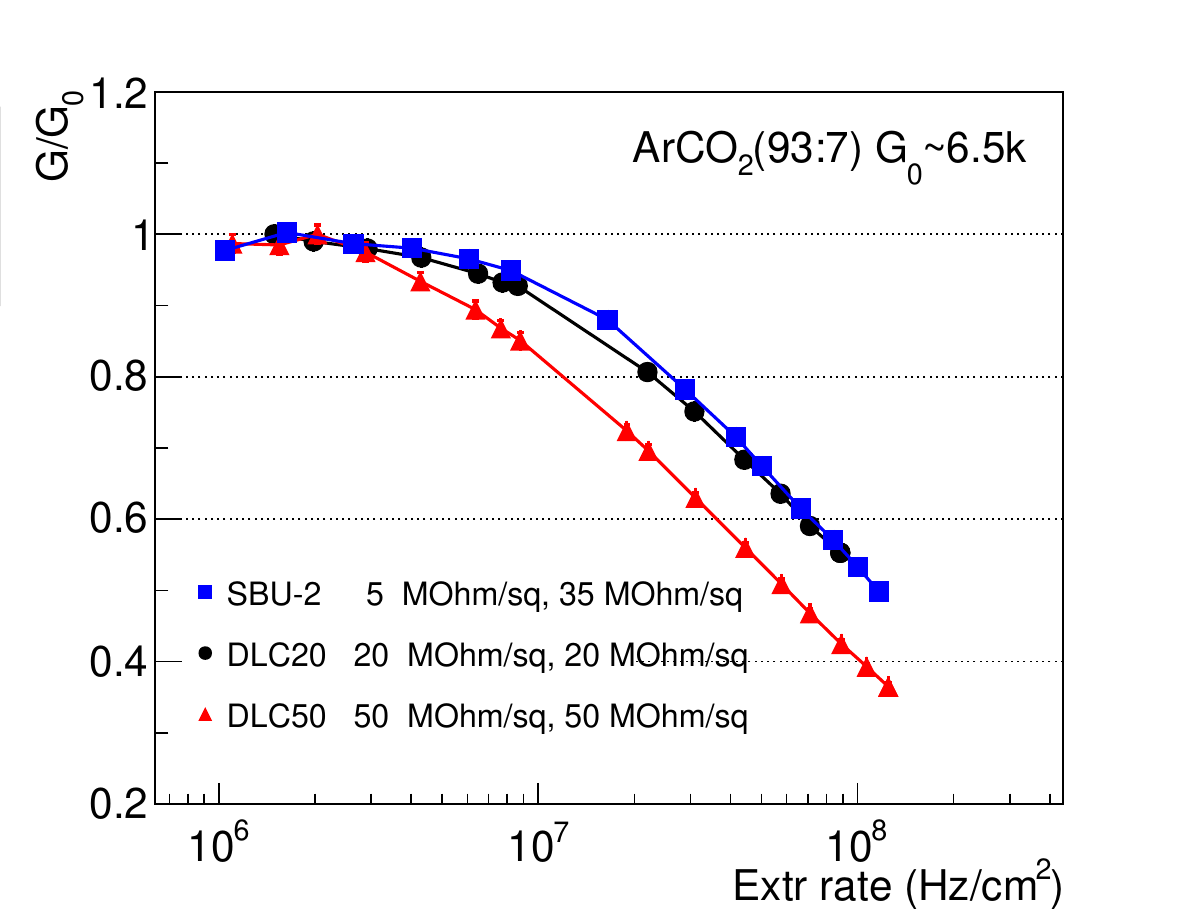}
\caption{Rate dependence of the gain normalised to the values at low rates G$_0$, measured in the DLC20 prototype  for different vias pitch (left) and in  detectors with different resistivities of the two DLC foils (right). 
\label{fig:ratecapability-pitch}}
\end{figure}

To investigate the impact of the grounding configuration of the double DLC layout
on the rate capability, we utilized the DLC20 detector, 
which features a network of grounding vias with pitches of 6 mm and 12 mm 
across two halves of the active area. In figure~\ref{fig:ratecapability-pitch}--left, we compare the performance at a gain of 10$^4$.
Both configurations exhibit no significant dependence on rates up to 1 MHz/cm$^2$. 
However, for higher rates, the gain begins to decrease. As expected, the gain reduction owing to voltage drop is more pronounced with for the charge evacuation vias with the larger pitch. Quantitatively, there is a difference of approximately 10\% 
between the two configurations at 10 MHz/cm$^2$.

Furthermore, we analysed the rate capability as a function of the surface resistivity 
of the two DLC foils. 
This comparison involved detectors with identical grounding evacuation network configurations
(with 6 mm pitch vias) and DLC resistivities of approximately 20 M$\Omega/\square$ on each foil (DLC20), 5 M$\Omega/\square$ and 35M$\Omega/\square$ (SBU-2) and 50 M$\Omega/\square$ on each foil (DLC50), all operated at a gain of about 6500. In figure~\ref{fig:ratecapability-pitch}--right, we observe consistent behaviour 
across all detectors up to approximately 5 MHz/cm$^2$, 
followed by a drop as expected,  more pronounced for the highest resistivity, amounting to about 10\% at 10 MHz/cm$^2$.

\subsubsection{Rate capability and ion backflow for the 20 M$\Omega/\square$ double DLC layer layout}
From the results presented so far, an optimal layout has been identified by adopting a double-layer 
of DLC foils each having approximately 20 M$\Omega/\square$ surface resistivity, 
with a network of dot grounding vias with a pitch of about 6 mm. 

\vspace{1.2\baselineskip}
{\noindent \bf Rate capability}

\begin{figure}[htbp] 
\centering
\includegraphics[width=.7\textwidth]{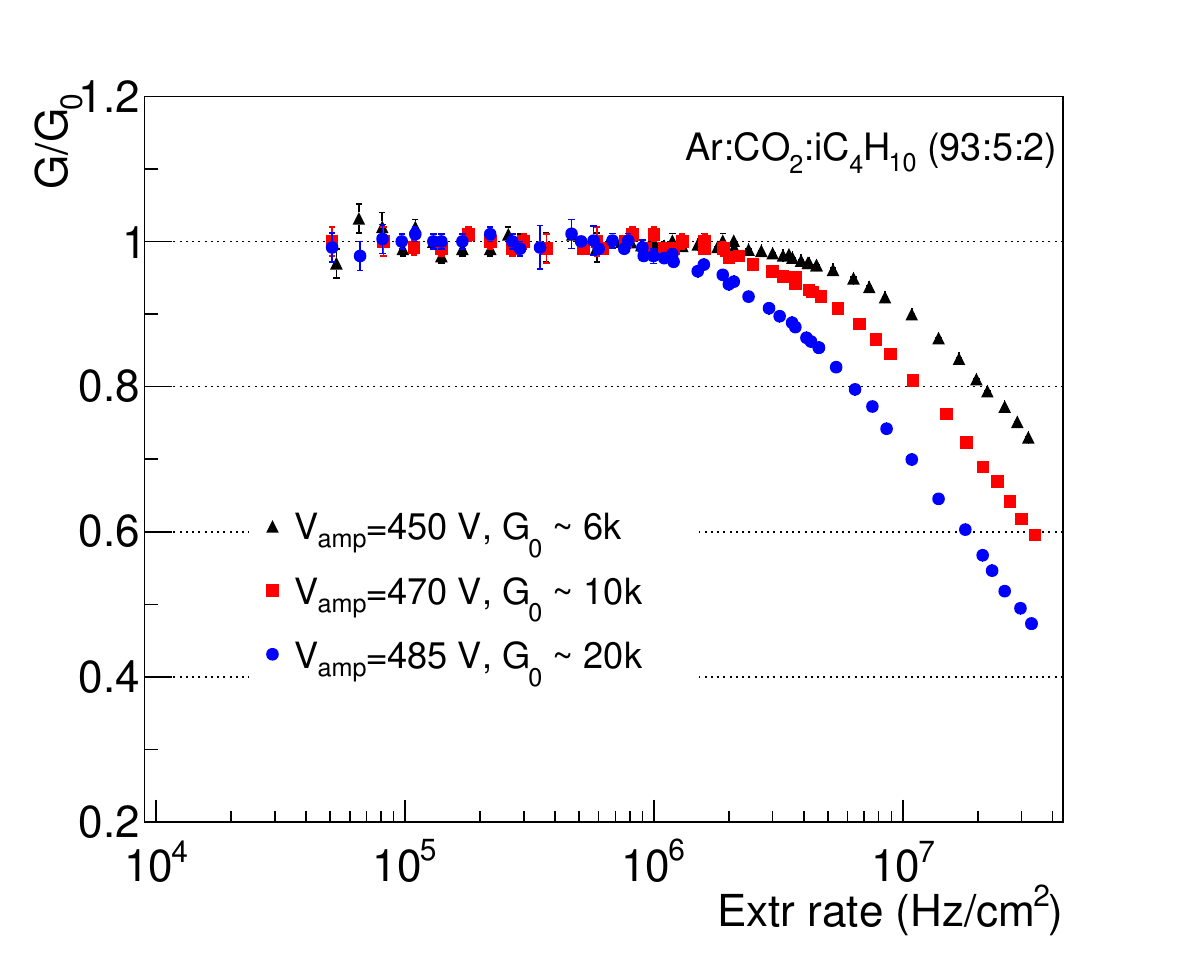}
\caption{Dependence of the gain normalised to the values at low rates G$_0$ = \num{6e3}, 10$^4$ and \num{2e4},
 measured on the DLC20 detector, operated with 
Ar--CO$_2$--C$_4$H$_{10}$ (93--5--2),
under variable flux irradiation from a  
Cu – X-ray generator (with 8.09 keV characteristic peak). 
\label{fig:ratecapability-dlc20}}
\end{figure}

The rate capability (gain versus hit rate) was measured with the detector DLC20 for different values 
of the amplification voltage, at fixed drift voltage (300 V, corresponding to E$_{\rm drift}$ = 600 V/cm) 
with a gas mixture of Ar--CO$_2$--iC$_4$H$_{10}$ (93--5--2).
In figure~\ref{fig:ratecapability-dlc20}, the rate capability is reported for gains of \num{6e3}, 10$^4$ and \num{2e4} 
as a function of the irradiation density of X-rays from the copper anode X--Ray generator. 
It is remarkable to note that the gain drop at 10 MHz/cm$^2$ is limited to 10\%, 
 for operational gains of G$_0$ = \num{6e3}, which is a typical gain that can be used in 
particle tracking detectors, and to 20\% and 30\%  for gains of 10$^4$ and \num{2e4}, respectively.
It must be also noted that these limited drops in the gain could be compensated (if needed) by an increase 
of few Volts of the amplification voltage.

\vspace{1.2\baselineskip}
{\noindent \bf Ion backflow}

From the same dataset, the ion backflow (IBF) was evaluated. 
The IBF is defined as the fraction of the average number of ions produced 
in the electron avalanche and the average number of the ions backflowing 
into the drift region. 
From the measurements of the currents in the drift and in the amplification circuits,
the IBF fraction can be estimated from the ratio of the current measured in the 
drift gap and the total current due to the multiplication processes: 
\begin{equation}
   {\rm IBF} = \rm I_{\rm drift}/(I_{\rm amp} + I_{\rm drift})
   \label{eq:IBF}
\end{equation}

\begin{figure}[htbp] 
\centering
\includegraphics[width=.7\textwidth]{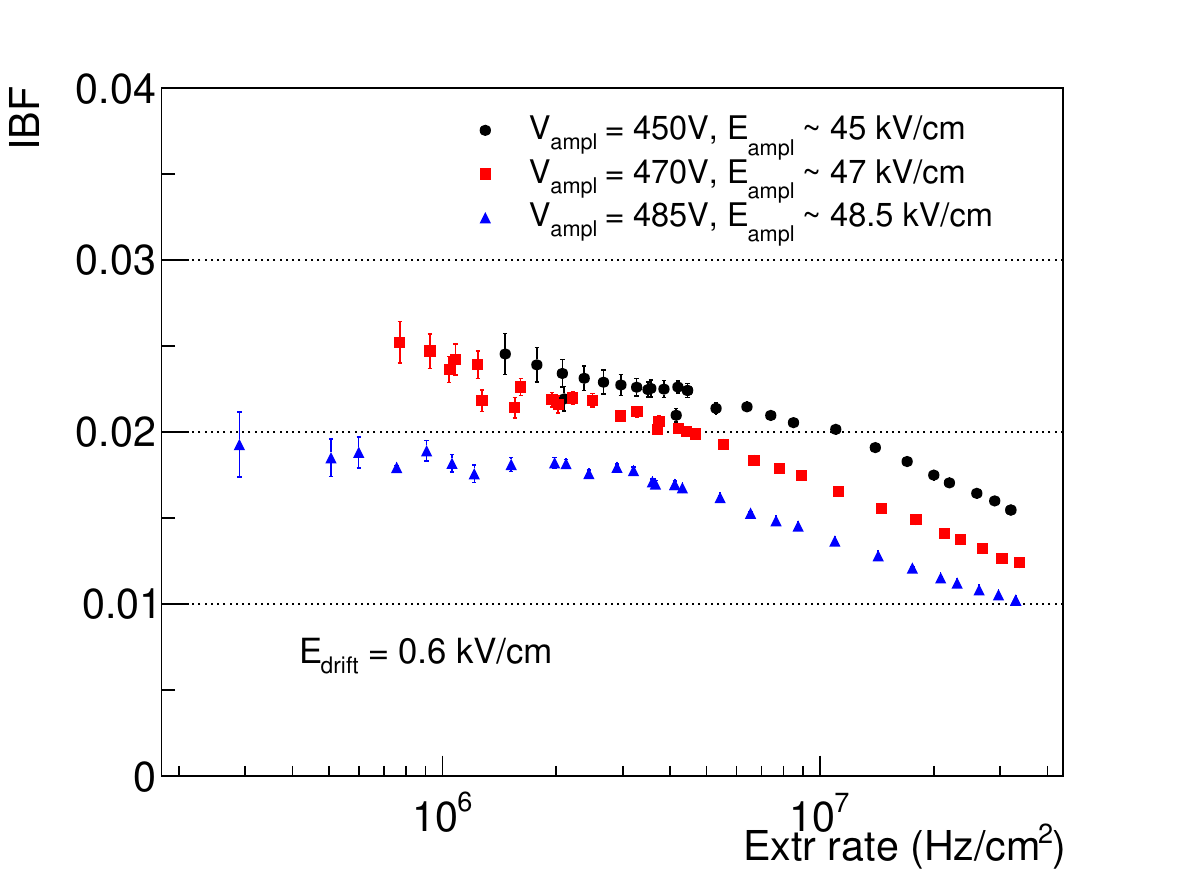}
\caption{Ion Backflow measured on the DLC20 detector, operated with 
Ar--CO$_2$--iC$_4$H$_{10}$ (93--5--2) at gains of \num{6e3}, 10$^4$ and \num{2e4}
(E$_{amp} \sim 45, 47, 48.5$ kV/cm, respectively)
under variable flux irradiation from a  
Cu – X-ray generator.
\label{fig:IBF}}
\end{figure}
In equation~\ref{eq:IBF}, the contribution of the primary 
ionisation current to I$_{\rm drift}$ is negligible.
In figure~\ref{fig:IBF}, the IBF as a function of the converted X-ray rate 
is reported at fixed drift voltage (E$_{\rm drift}$ = 600 V/cm), for different values of the amplification voltage. 
As reported in~\cite{IBF-Colas, IBF-Chefdeville, IBF-Bhattacharya} 
the IBF is roughly proportional to the ratio of the drift and amplification electric fields.
It is therefore expected to decrease from the lowest amplification field ($\sim$45 kV/cm) 
to its highest value ($\sim$48 kV/cm). 
We also note a decrease as a function of the rate, simulation studies are needed to further investigate this effect. 
The measured IBF fraction in the range of 1--3 \% is a good result for a single stage amplification system. Moreover, this measurement is 
in good agreement with bulk 
Micromegas measured in similar conditions in~\cite{IBF-Bhattacharya}. 
It must be also stressed that the IBF also depends on the structure of the mesh and 
on the diffusion properties of the gas. 
An optimisation of the construction design and operation parameters 
could be therefore envisaged to further reduce the IBF fraction.

\subsection{Rate capability and dependence on the irradiated area for the medium-size detector}
\label{sec:dependenceonarea}

The analysis of the detector response, 
examining gain stability in relation to the irradiation rate, 
has been conducted on Paddy400, in the same way as outlined in section~\ref{sec:ratecapability_method}. 
In addition, a study on the dependence of the irradiated surface was carried out.

\begin{figure}[htbp]
\centering
\includegraphics[width=.465\textwidth]{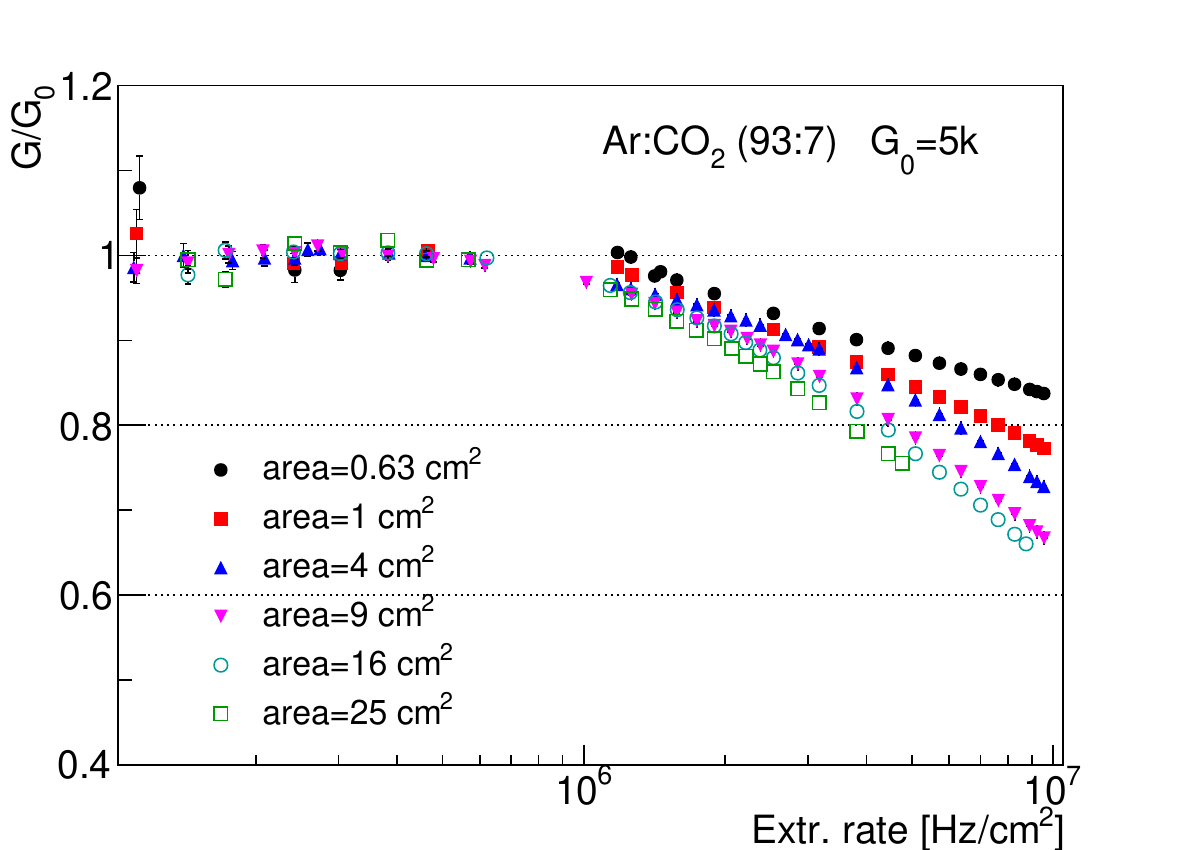}
\includegraphics[width=.49\textwidth]{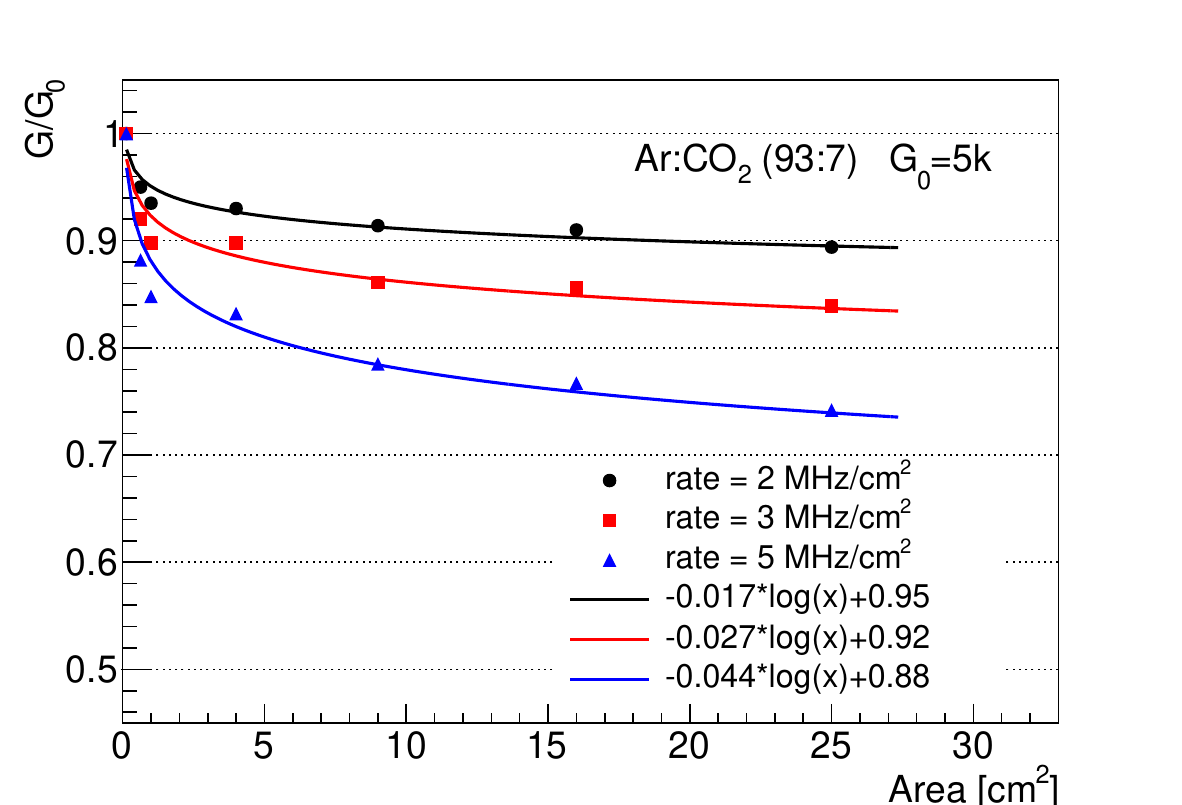}
\caption{Dependence of the gains of the Paddy400-1 detector, normalised to its value at low rates, 
for different uniformly irradiated surfaces. 
Left: as a function of hit rates; 
Right: as a function of the irradiated surface at three values of constant density rates: 2, 3, 5 MHz/cm$^2$. }
\label{fig:ratecapability-Paddy400}
\end{figure}

All measurements reported in section~\ref{sec:ratecapability}, 
were based on an X-rays exposed surface of 0.79 cm$^2$. 
To study the dependence of the detector’s response on the irradiated area, 
masks with different apertures have been used. 
In figure~\ref{fig:ratecapability-Paddy400}-left, the measurements of the gain drop
as a function of the X-Rays hit conversion rate is reported 
for 6 different irradiated surfaces in the range 0.63 – 25 cm$^2$. 
Accurate measurements were done to validate uniform flux density over the full exposed surface.
The detector was operated with Ar--CO$_2$ (93--7) gas mixture at a gain of \num{5e3}. 
The missing data points in the range 0.7--1 MHz/cm$^2$ are unfortunately due to
instabilities in the range of $\sim$0.2--0.4 mA of the X-Rays generator.
However, as shown in figure~\ref{fig:xraycalib}, its stability and linearity outside that region was extensively validated.
The data points for the apertures at 16 and 25 cm$^2$ could not extend to the maximum rates 
since the current limit of \SI{20}{\micro\ampere}  of the HV power supply was reached. 
The gain drop as a function of the irradiated surface is reported in 
figure~\ref{fig:ratecapability-Paddy400}-right for three values of the hit rates: 
at 2, 3 and 5 MHz/cm$^2$ (all corresponding to an ionisation from MIP rates well above 10 MHz/cm$^2$). 

Even if the network of grounding vias is intended to establish independent ``cells'' 
(each 8x8 mm$^2$ in size) and limit the gain drop for constant-density irradiation on 
surfaces larger than the unit cell, it can be observed that there is still a cumulative 
effect of charging-up on the resistive layer. 
This effect results in a trend of increasing gain drop with larger surfaces. 
These trends, characterised by a slow progression, 
have been empirically fitted with a logarithmic dependence.
If extrapolated to an area  of \numproduct{50x50} cm$^2$ 
(a reasonable size for a typical detector for large area future experiments), 
fully irradiated with a flux of 3 MHz/cm$^2$ of (highly ionising) particles, 
the total drop would be around 30\%, an effect that could be also easily compensated 
by an increase of about 10 Volts of the amplification voltage  (see figure~\ref{fig:fig-gain}-left). 

For a comprehensive overview, it is useful to recall the findings also related to the pad-patterned 
prototypes (PAD-P types). 
In these particular configurations, all pads are entirely independent and 
insulated from each other. 
Consequently, the detector's response is expected to remain unaffected 
by the irradiated surface (for areas exceeding the size of an individual pad), 
given a constant density rate. 
This consideration aligns with the results documented in~\cite{INSTR2020}, 
extending up to 40 MHz/cm$^2$, 
for surfaces within the range 0.79--9 cm$^2$.


The objective of future measurements is to broaden the study to encompass larger surfaces. 
However, finding a facility with a readily available source of variable and high-intensity 
radiation across large surfaces, while maintaining a consistently uniform density throughout the entire area, 
proves to be a challenging task.


\section{Test-beam results from the medium-size detector}
\label{sec:Paddy400-testbeam}

In 2022 and 2023 the detectors were tested with 150 GeV muon beams in the H4 SPS Test-beam area at CERN. 
A frame containing two scintillators for trigger, two X-Y strip resistive Micromegas providing external tracking (TMM), 
the two Paddy400 and two small-size pixelized double DLC Micromegas, were used. 
The two Paddy400 detectors were assembled in the setup in the ``front-to-front'' configuration, 
as shown in figure~\ref{fig:Paddy400-layout2}. The frame hosting the chambers was constructed in such a way that the detector under test could be inclined by an angle between 0$^{\rm{o}}$ and 40$^{\rm{o}}$ with respect to the beam direction. 
In this section, a selection of the main results obtained for Paddy400-1 and Paddy400-2 detectors are reported.  
The chambers were operated with gas mixtures of Ar--CO$_2$--iC$_4$H$_{10}$ (93--5--2)
or Ar--CF$_4$--iC$_4$H$_{10}$ (88--10--2), sharing the same cathode plane,  
and were readout using the APV25 \cite{apv} front-end ASICs connected to the Scalable Readout System \cite{srs}.

\vspace{1.2\baselineskip}
{\noindent \bf Efficiency}

\begin{figure}[htbp] 
\centering
\includegraphics[width=.48\textwidth]{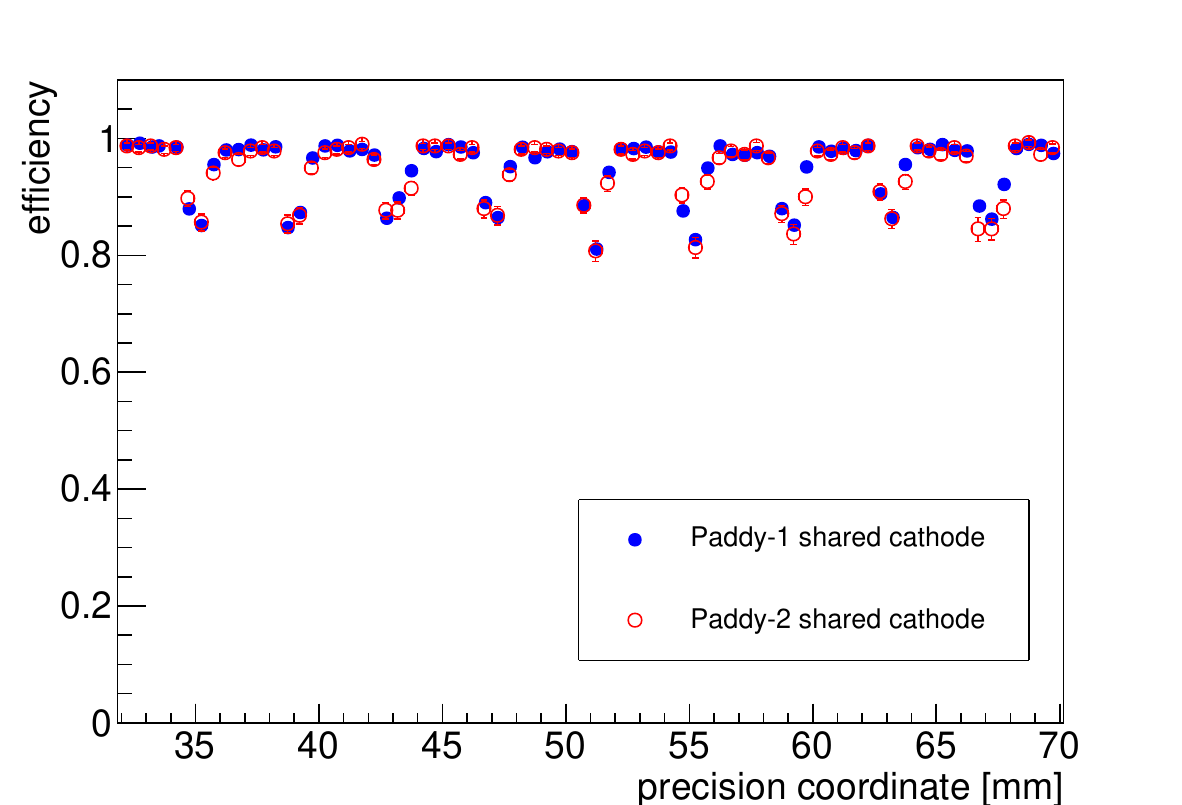}
\includegraphics[width=.48\textwidth]{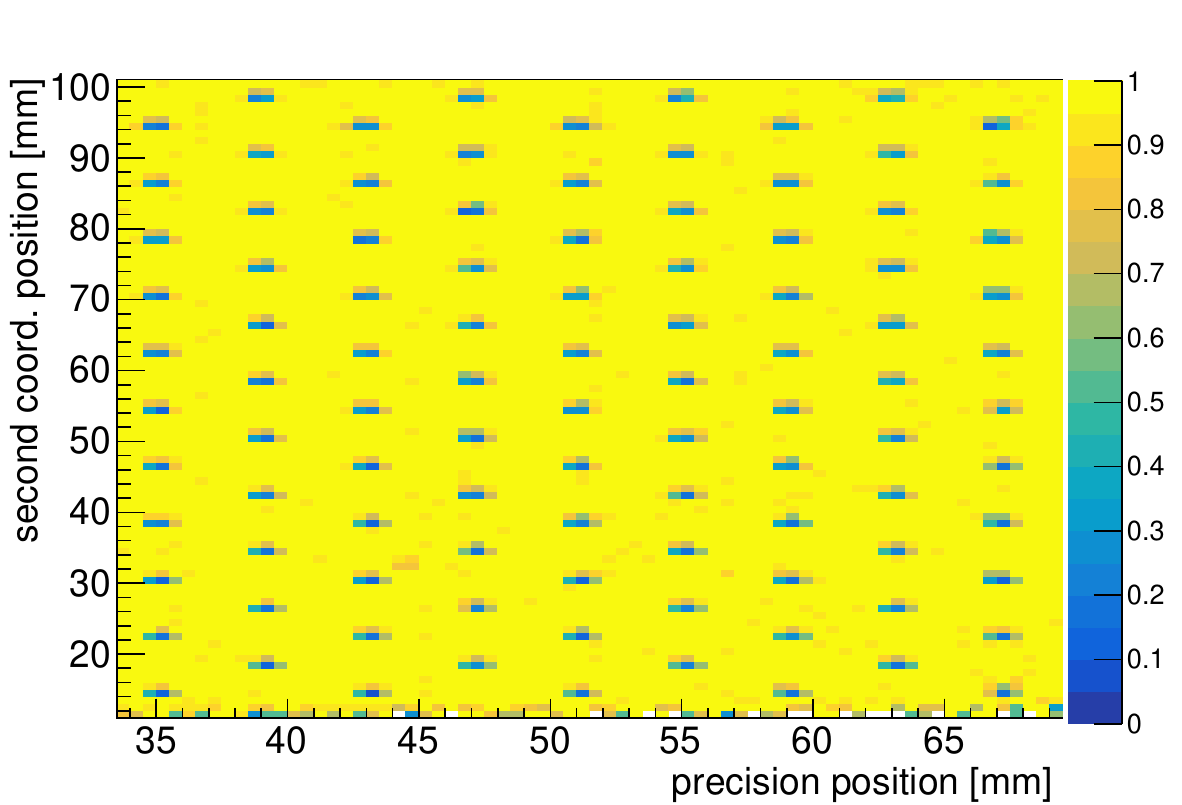}
\caption{Efficiencies for both Paddy400-1 and Paddy-400-2 as a function of the precision coordinate. 
The uncertainties include only statistical source (left). The 2D map of the x-y plane efficiency for Paddy400-1 (right).}
\label{fig:Fig_paddy400Efficiency}
\end{figure}

In order to evaluate the chamber efficiencies, tracks are selected requiring single clusters in the two TMM chambers 
in both x and y coordinates, and the cluster positions, for all chambers readout planes, are evaluated with the 
standard charge-weighted centroid. 
The track trajectories are obtained by fitting the reconstructed x and y cluster positions in the TMM detectors. 
In this way the extrapolated coordinate, ${\rm x_{trk}}$, at the positions of the chambers under test can be computed. 
The chamber misalignments, both translations  and rotations, are corrected by analysing the  average values of the residuals with respect to the extrapolated track positions.
The detection efficiency is defined as the ratio between the number of cases where a cluster in the Paddy400 chamber 
under study is found within a fiducial range of 1.5 mm around the extrapolated position and the number of extrapolated tracks.
The chambers were operated at an amplification voltage of 460V, 
corresponding to a gain of about \num{8e3} (in Ar--CF$_4$--iC$_4$H$_{10}$).
The efficiency along the precision coordinate and the two-dimensional efficiency are shown in figure~\ref{fig:Fig_paddy400Efficiency}  
for perpendicular tracks. A drop of the efficiency, to about 20$\%$, is observed every 4 mm (8 mm) along the precision (second) coordinate, 
which corresponds to the pillar locations, while it is very close to 100$\%$ elsewhere. 
It's worth noting that inefficiency is more pronounced in the chambers under test 
since they have a relatively large pillar diameter, of 0.7~mm. 
The efficiency can be optimised by reducing the pillar diameters, using the SBU technique (see section~\ref{sec:sbu}), 
which can reduce the area covered by the pillars by a factor of about 5.
In average, an efficiency close to 98\% is achieved.
For not perpendicular tracks, the effect of  pillars is expected to be reduced, since the track inclination 
is likely to result in ionization across areas larger than the pillar dimensions.

\vspace{1.2\baselineskip}
{\noindent \bf Spatial Resolution}

The spatial resolution of the Paddy400 detectors is determined using  perpendicular tracks, by measuring the width of the distribution of 
cluster position residuals with respect to the extrapolated position of the tracks reconstructed from the TMM chambers. The extrapolation uncertainties, of about \SI{50}{\um}, is subtracted from the measurements. 
Only perpendicular tracks are used in this analysis.
The cluster position is evaluated with an extended definition of the charge weighted centroid:
\begin{equation}
 {  x_c=  \frac{\sum x_i q_i^p}{\sum q_i^p}}
\end{equation}
where $\rm x_i$ and $\rm q_i$ are the position and the charge of the i-th strip of the cluster in the precision coordinate, 
respectively. The parameter p, commonly set to 1, is instead extracted by minimising  the residuals 
\begin{equation}
 {  \Delta x = x_{trk} - \frac{\sum x_i q_i^p}{\sum q_i^p}}
\end{equation}

In the centroid reweighting optimisation procedure
a double-gaussian fit function to the distribution of residuals is performed and 
the optimal parameter p is the one minimising its width.

\begin{figure}[htbp] 
\centering
\includegraphics[width=.49\textwidth]{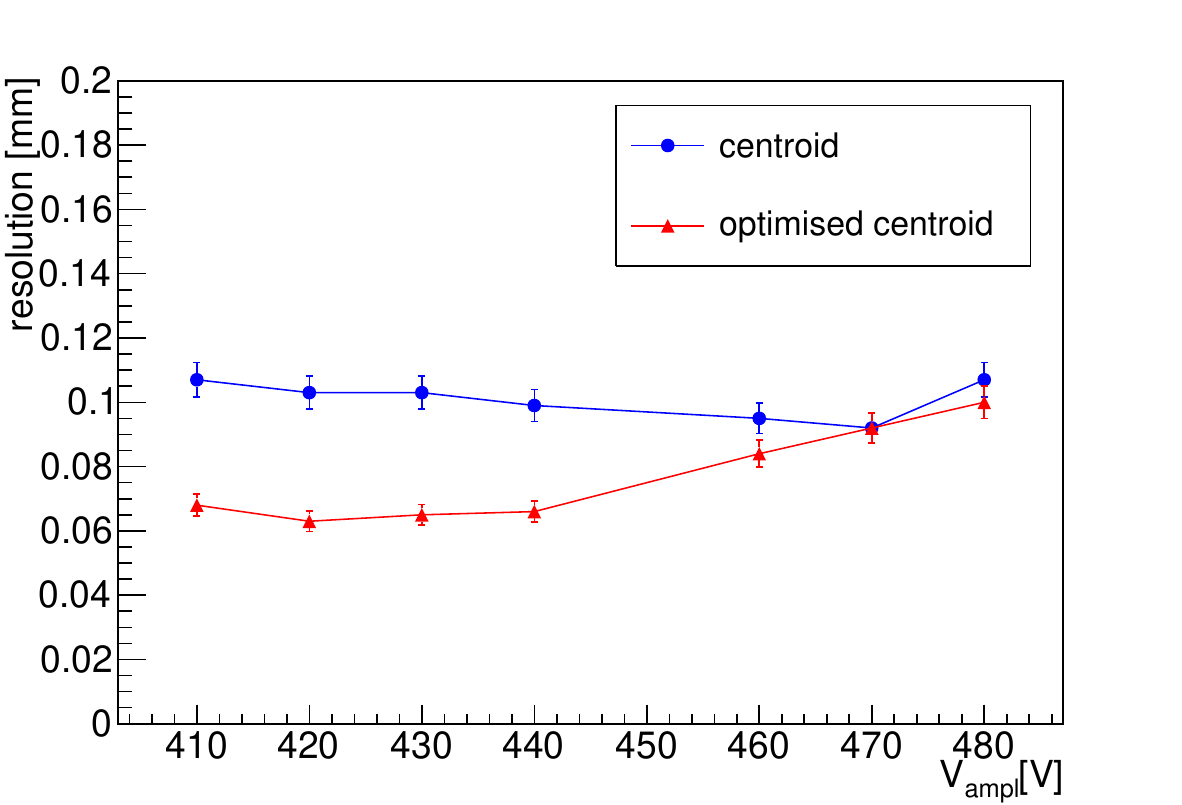}
\includegraphics[width=.47\textwidth]{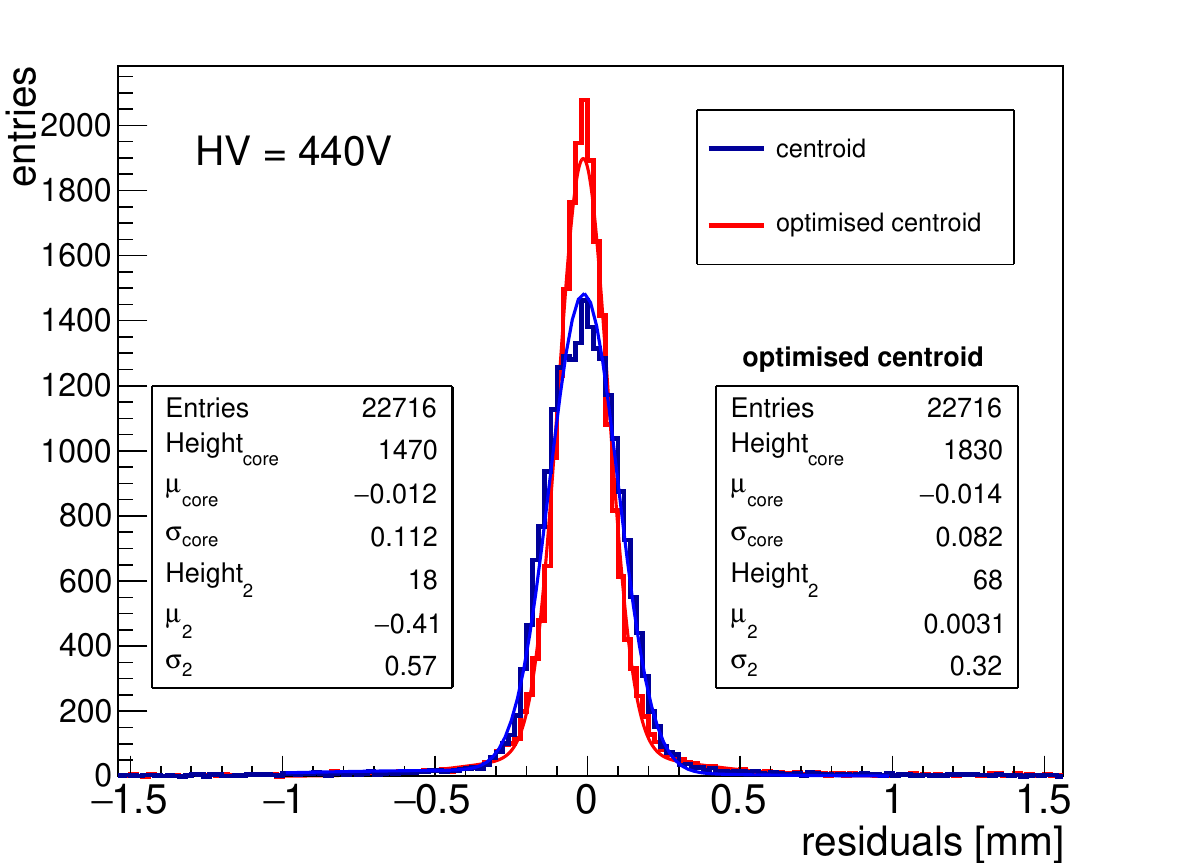}
\caption{Paddy400-1 spatial resolution, from the precision coordinate, as a function of the amplification voltage (left) for not optimised centroid method (blue dots) and after the centroid optimisation (red triangles).  The uncertainties include both statistical and systematic sources. On the right side, the distribution of the residuals for HV=440 V before and after the centroid optimisation is shown together with the results from a double-Gaussian fit.}
\label{fig:paddy400_resol}
\end{figure}

The results obtained for the detector Paddy400-1 is shown in figure \ref{fig:paddy400_resol}-left as a function of the amplification voltage. The resolution is measured as the $\sigma$ of the core Gaussian in a double-gaussian fit. 
The optimised centroid improves the spatial resolution up to 440~V of about 35$\%$ and  
resolutions better than \SI{70}{\um} can be obtained for perpendicular tracks from the precision coordinate. 
The optimised value for p is found to be 0.65. 
For higher values of amplification voltages, higher values of p, around 1, are found and the improvement 
from the optimisation is marginal. This has been tracked down to poor measurements of the charge with APV25 at 
high gain due to saturation. 
In this regime the charge weighted centroid method is less accurate. 
The statistical uncertainty on the resolution is negligible, while the systematic
uncertainty, mostly due to the fit procedure, 
is estimated to be  5$\%$.
In figure \ref{fig:paddy400_resol}-right, the distribution of the residuals for an amplification 
voltage of 440 V is shown for the standard centroid method ($p=1$) and the optimised centroid ($p=0.65$). The parameters of the double-Gaussian fit are also shown: the distribution is almost entirely described by the core Gaussian function. 
The results are similar for both Paddy400-1 and Paddy400-2.

\vspace{1.2\baselineskip}
{\noindent \bf Time Resolution}

\begin{figure}[htbp] 
\centering
\includegraphics[width=.44\textwidth]{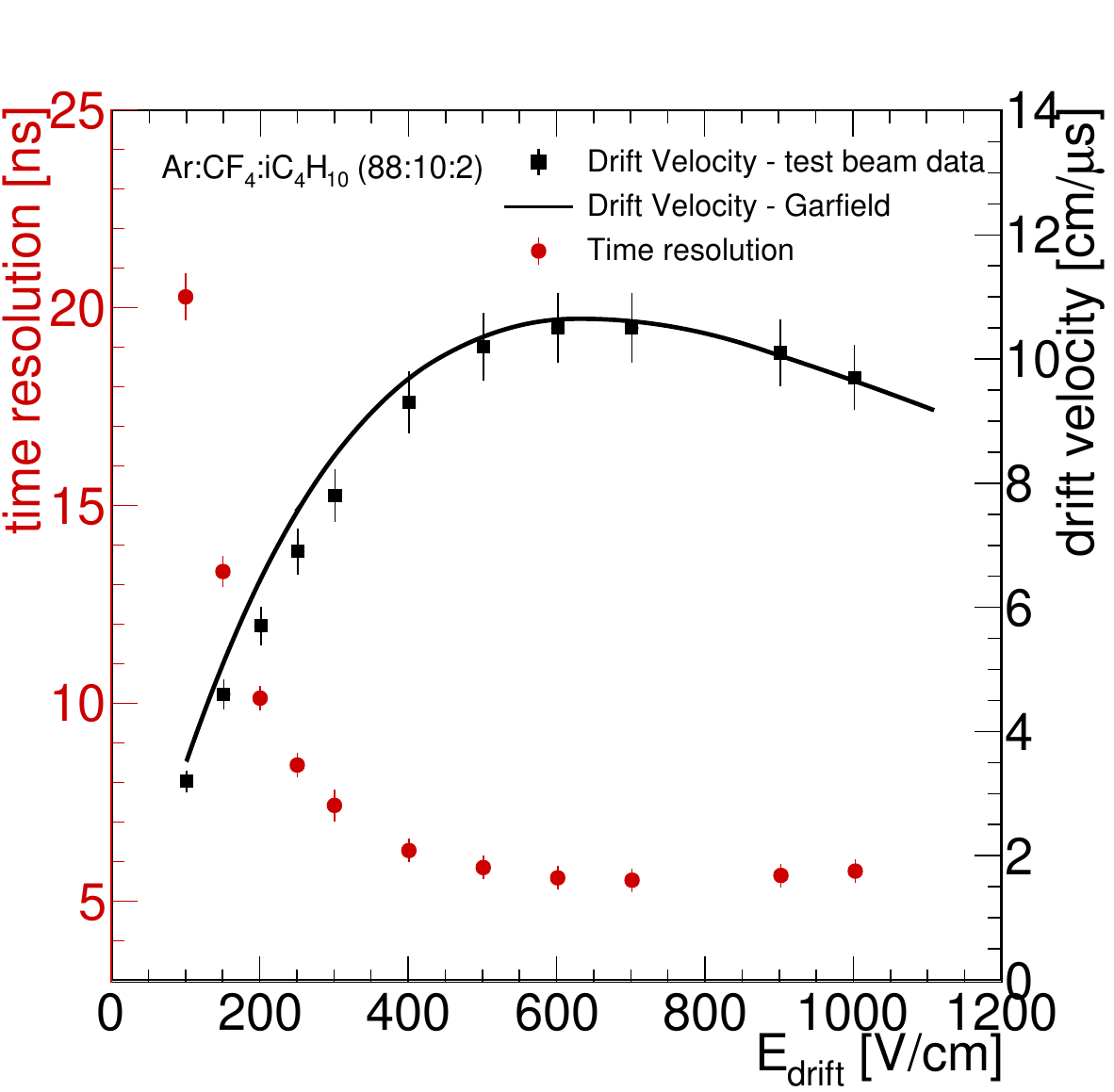}
\hspace{5mm}
\includegraphics[width=.43\textwidth]{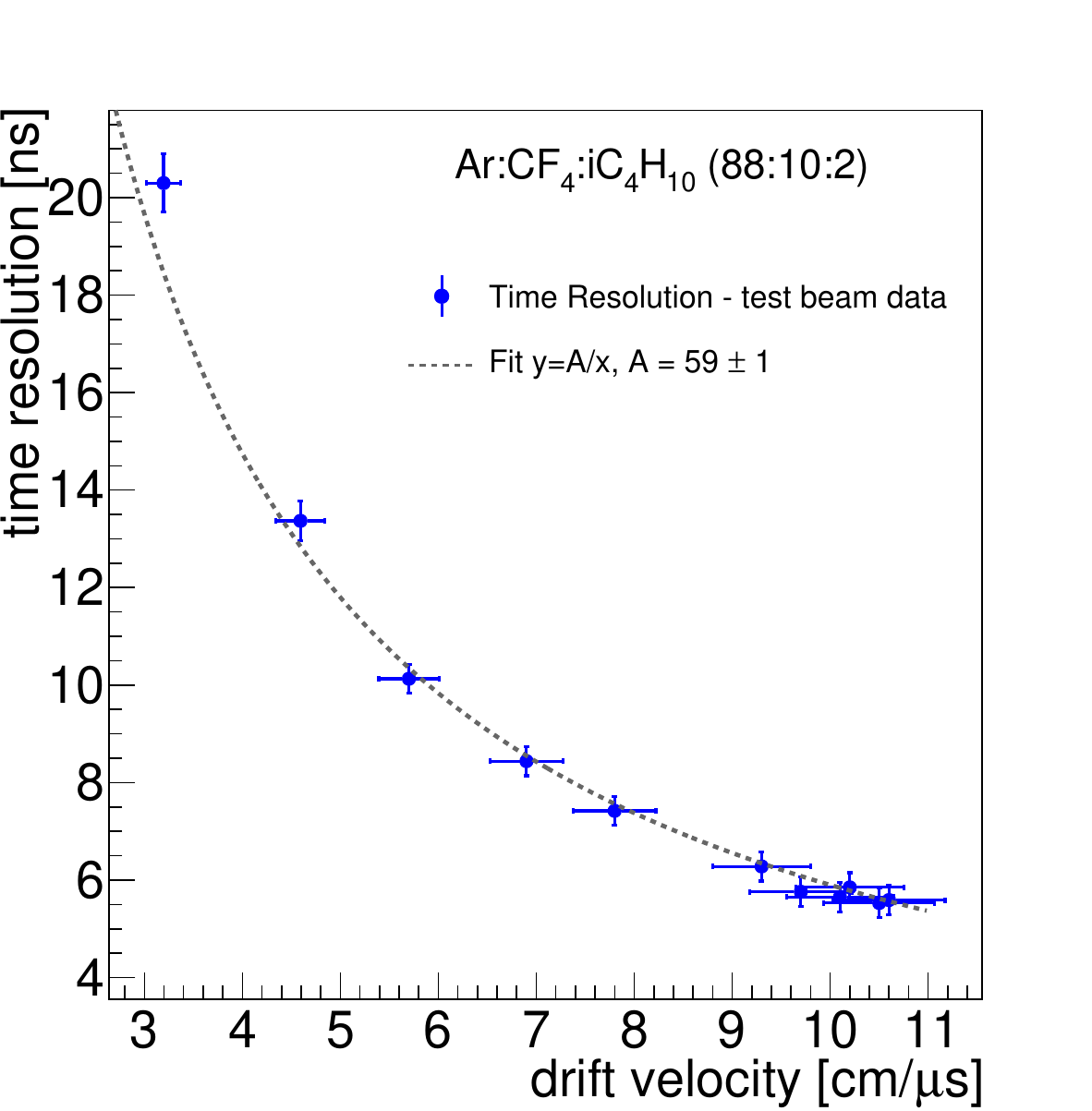}
\caption{Right: time resolution and measured drift velocity  as a function of the drift electric field. Drift velocity from Garfield simulation is also reported. Left: time resolution as a function of the measured drift velocity. 
\label{fig:Paddy400-resolutions}}
\end{figure}

\begin{figure}[htbp!] 
\centering
\includegraphics[width=.45\textwidth]{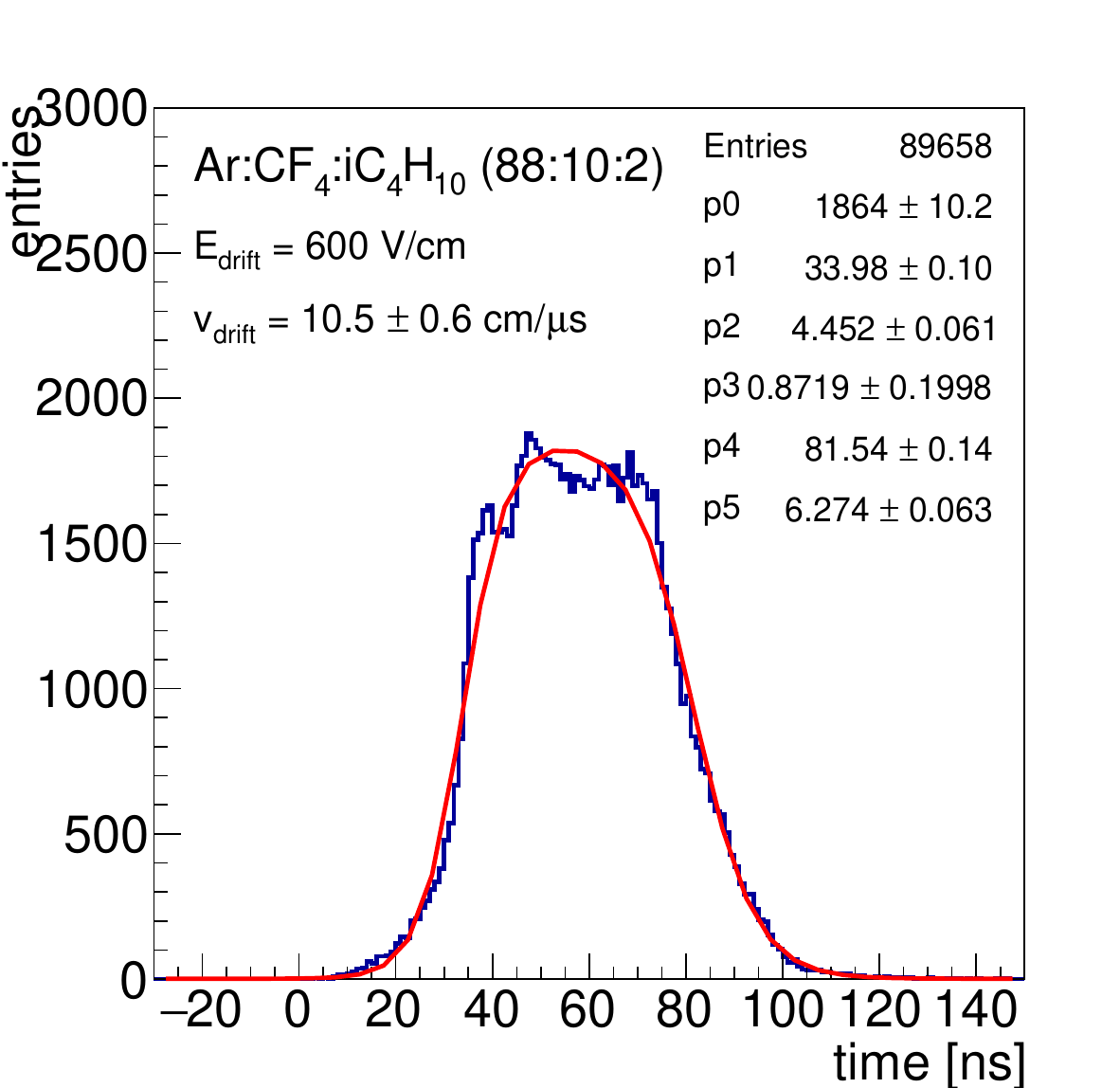}
\includegraphics[width=.45\textwidth]{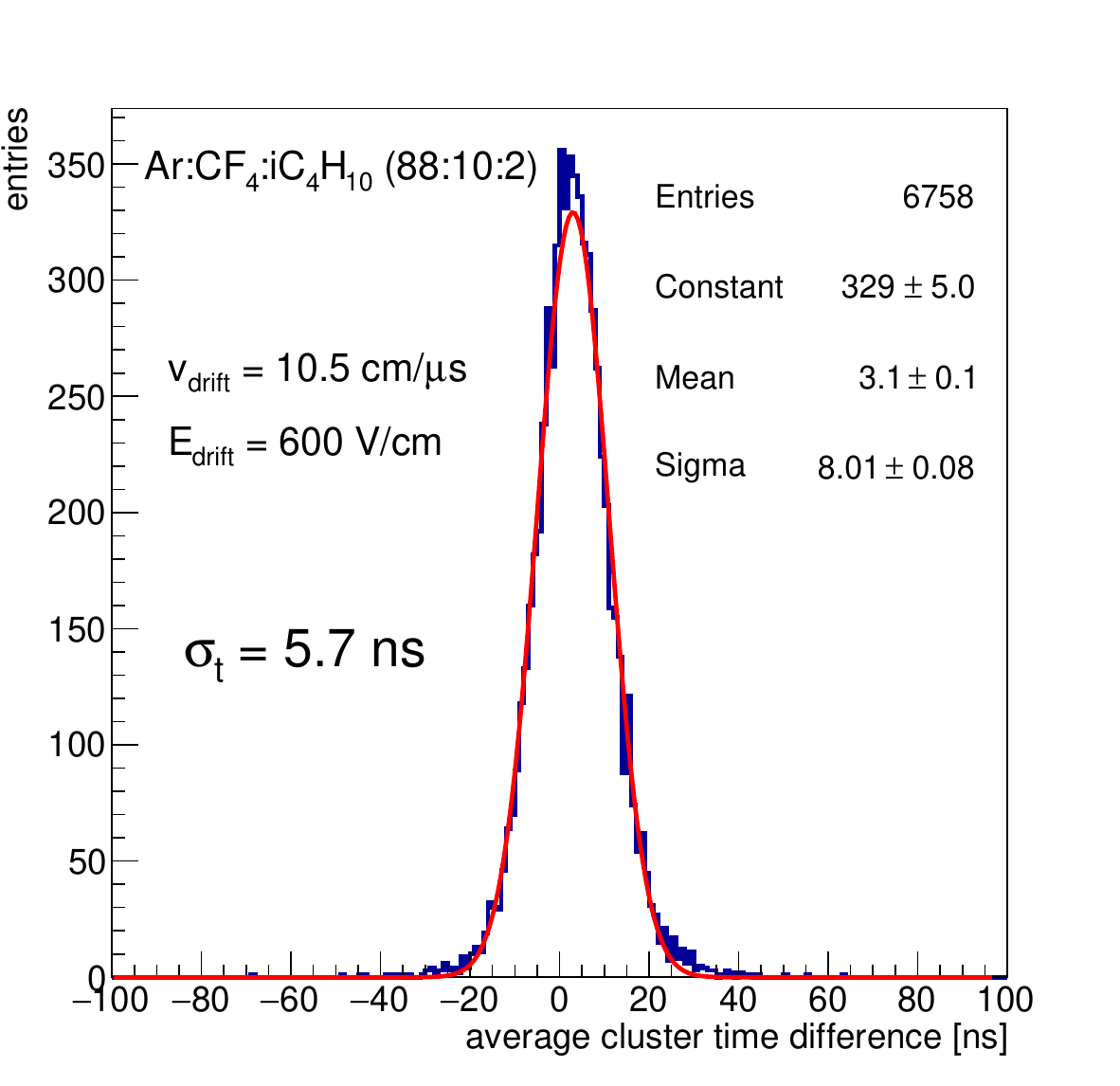}
\caption{Inclusive time distribution for a Paddy-400 chamber obtained using test beam data with the chamber inclined by 35$^{\rm{o}}$ with respect to the beam line (left). Average cluster time differences of two Paddy400 chambers for the higher drift velocity (right).
\label{fig:Paddy400-timebox}}
\end{figure}

The time resolution of the Paddy400 chambers was also measured, 
%
with the 
gas mixture Ar--CF4--iC$_4$H$_{10}$ (88--10--2), that provides an electron drift velocity higher than in other gases,
 and with different electric fields applied to the drift gap. For each pad above threshold, the APV25 returns the shape of the charge sensitive amplifier with a sampling in 25 ns bins. The pad time is then extracted by fitting the rise of the APV25 signal with a Fermi-Dirac function~\cite{Alexopoulos-first}.
To avoid a worsening of the measurements from the superposition of signals from a larger number of primary ionisations taking place above a single pad, the measurements described in the following were carried out with  chambers inclined by 35$^{\rm{o}}$ with respect to the beam direction. 
The cluster time is defined as the average time of all pads belonging to the cluster. 
Other definitions, like the time of the earliest hit, were tested, giving slightly worse results. 
The time resolution is evaluated from a Gaussian fit to the distribution of the difference of cluster times between the two Paddy400 chambers crossed by the same track.  An uncertainty of about 3-4$\%$  on the time resolution has been estimated from the fit procedure. 
After ensuring that the time response is similar for both devices under test, the final reported resolution is  divided by a factor of $\sqrt 2$.
The jitter of 25 ns  coming from  APV25 is cancelled in the subtraction of the cluster times, since the clock is synchronous for all APV25 in the setup. 
In figure \ref{fig:Paddy400-resolutions}, the time resolution as a function of  the drift electric field (left)  and of the drift velocity (right) is shown. 
The drift velocity has been measured in each chamber using the time distribution of all hits in  reconstructed clusters, from which the time interval to cross the drift gap has been evaluated. A representative time distribution is shown in figure~\ref{fig:Paddy400-timebox}-left. 
These distributions are fitted with a double-side Fermi-Dirac function that allows  to get the minimum time (t$_{\rm{min}}$) and maximum time (t$_{\rm{max}}$) as those corresponding to the inflection points of the rising and trailing edges of the time spectrum. The drift velocity is then  measured as 
\begin{equation}
 {\rm  v_d = \frac{D}{t_{max}-t_{min}}}
\end{equation}
where D is the drift gap size (5 mm).
The uncertainties on the drift velocity come from the uncertainty on the drift gap size (about $2\%$) and from the fit procedure (about $5\%$). 
The distribution of the average cluster time differences of the two Paddy400 chambers for the maximum value of the drift velocity of 10.5 cm/\SI{}{\micro\second}  is shown in figure \ref{fig:Paddy400-timebox}-right. 
From a Gaussian fit a time resolution of 5.7 ns is measured. 
It is important to consider that the measured resolutions include a convolution of electronics effects.
Based on preliminary estimates, this contribution amounts to a few ns.

\section{Summary and future prospects}
\label{sec:summary}

This paper reports on the development and performance of resistive small pads  Micromegas detectors for next-generation particle physics  applications.
Different spark protection resistive layouts have been implemented on several small-pad Micromegas prototypes.
The optimal detector configuration employs a resistive layout based on a double-layer of uniform DLC resistive foils, with a network of dot
patterned vias for fast charge evacuation. 
The performances achieved include stable operation with gains greater than 10$^4$ and rate capability  up to 10 MHz/cm$^2$, 
also with large area irradiation. The wide range of stability allows to define working points with large margin and flexibility. 

From test-beam data, an average efficiency close to 98$\%$ and a spatial resolution 
for perpendicular tracks below \SI{70}{\um} have been measured.
Measurements of time resolutions are reported in a wide range of electron 
drift velocity, reaching a value below 6 ns, including the electronic contribution.

Future developments will address the detector scalability towards large area for large systems for future experiments.
A larger pixelized prototype (50 × 40 cm$^2$ wide) has been already successfully realised and it is currently under test.
Another goal is the simplification of the construction process, also in view of technological transfer to industry. 
A dedicated R$\&$D is currently ongoing within industry, where several  prototypes have been already produced.
For low rate applications, the possibility of use of a single DLC foil with the charge evacuation from the side will be explored.  
Finally, the capacitive sharing technique~\cite{kondo} is being investigated and applied to resistive pixelised Micromegas, 
allowing a reduction of readout channels while maintaining a good spatial resolution.


\acknowledgments

We would like to thank the CERN MPT workshop, in particular R. De Oliveira and his group, for ideas, 
discussions and the construction of the detectors; 
E. Oliveri and the whole RD51 Collaboration for support with the tests at the Gas Detector Development (GDD) Laboratory 
and for the test-beam at CERN.





\end{document}